\documentclass[aps,prd,reprint,groupedaddress]{revtex4-2}
\usepackage{amsfonts}
\usepackage{amsmath}
\usepackage{amssymb}
\usepackage{appendix}
\usepackage{array}
\usepackage{bm}
\usepackage{bbm}

\usepackage{graphicx}
\usepackage{caption}
\usepackage{cases}
\usepackage{color}
\usepackage{xcolor}
\definecolor{deepblue}{rgb}{0,0.45,0.8}
\usepackage[colorlinks,linktocpage,linkcolor=blue,citecolor=cyan]{hyperref}

\usepackage{subfigure}

\def\nn{\nonumber}

\begin{document}

\title{Detecting gravitational wave background with equivalent configurations in the network of space based optical lattice clocks}

\author{Mingzhi Lou$^{1,4,5}$}
\email{loumingzhi23@mails.ucas.ac.cn}

\author{Hong Su$^{3}$}

\author{Tao Yang$^{3}$}

\author{Yun-Long Zhang$^{2,1}$}
\email{zhangyunlong@nao.cas.cn}

\affiliation{$^{1}$School of Fundamental Physics and Mathematical Sciences, Hangzhou Institute for Advanced Study, UCAS, Hangzhou 310024, China}

\affiliation{$^{2}$National Astronomical Observatories, Chinese Academy of Sciences,
Beijing 100101, China}

\affiliation{$^{3}$School of Physics and Technology, Wuhan University, Wuhan 430072, China}

\affiliation{$^{4}$CAS Key Laboratory of Theoretical Physics, Institute of Theoretical Physics, Chinese Academy of Sciences, Beijing 100190, China} 

\affiliation{$^{5}$Taiji Laboratory for Gravitational Wave Universe (Beijing/Hangzhou), University of Chinese Academy of Sciences, Beijing 100049, China}

\begin{abstract}
This paper studies the use of optical lattice clock (OLC) detector networks for detecting the stochastic gravitational-wave background (SGWB). Starting from the cross-correlation formalism for two OLC detectors, we analyze how the detector geometry influences the overlap reduction function (ORF) and systematically search for configuration transformations that preserve the modulus of the ORF. We identify an equivalent transformation in which the emitting and receiving ends of both OLC links are exchanged, while the modulus of the ORF remains invariant. We then numerically compare the ORFs of isosceles trapezoidal configurations with different separations and included angles. Based on these results, we design a feasible four-spacecraft orbital configuration and evaluate its strain sensitivity and noise energy-density spectrum in comparison with LISA, Taiji, and TianQin. 
\end{abstract}

\maketitle
\tableofcontents
\allowdisplaybreaks

\section{Introduction}

Since the first direct observation of the gravitational wave from the binary black hole merger by LIGO in 2015~\cite{LIGOScientific:2016aoc}, ground-based detectors  have detected dozens of compact binary coalescences, profoundly enhancing our understanding of compact objects like black holes and neutron stars~\cite{LIGOScientific:2017vwq, LIGOScientific:2016lio, LIGOScientific:2021qlt}. However, due to limitations from arm length and seismic noise in  LIGO~\cite{LIGOScientific:2014pky}, Virgo~\cite{VIRGO:2014yos}, and KAGRA~\cite{Somiya:2011np}, the sensitive band of terrestrial detectors is mainly above 10 Hz~\cite{Punturo:2010zz}. To detect low frequency gravitational waves in the millihertz range, space-based detectors are required. The LISA mission is one of the space-based laser interferometric gravitational-wave detectors, which will be sensitive in the 0.1 mHz to 0.1 Hz band~\cite{eLISA:2013xep}. Concurrently, the Taiji~\cite{Hu:2017mde} and TianQin~\cite{TianQin:2015yph} programs aim to explore the low-frequency gravitational-wave universe with similar architectures. In the even lower nanohertz frequency band, Pulsar Timing Arrays (PTAs)~\cite{Detweiler:1979wn} utilize the timing signals from an array of millisecond pulsars to detect GWs from the cosmic population of supermassive black hole binaries~\cite{Burke-Spolaor:2018bvk}, and have reported evidence for a background signal~\cite{NANOGrav:2020bcs,Goncharov:2021oub,EPTA:2021crs,Antoniadis:2022pcn}.

In recent years, gravitational-wave detection schemes based on optical lattice clocks (OLCs) have been proposed as a newly emerging technology, demonstrating great potential~\cite{He:2020elt,Ebisuzaki:2018ujm}. Indeed, OLCs possess exceptional stability and precision~\cite{Hinkley:2013oos,Bloom:2013uoa}. By comparing the frequency variations of optical signals between two spatially separated OLCs, it is possible to sense spacetime perturbations caused by passing gravitational waves~\cite{Kolkowitz:2016wyg}.
One of the core scientific objectives in gravitational wave cosmology~\cite{Christensen:2018iqi} is the stochastic gravitational wave background (SGWB). It may originate from physical processes in the early universe~\cite{Maggiore:1999vm,Caprini:2018mtu}, such as cosmic strings~\cite{Damour:2001bk,Siemens:2006yp}, first-order phase transitions~\cite{Witten:1984rs,Kamionkowski:1993fg}, and inflation~\cite{Mukhanov:1990me,Turner:1996ck}. It could also arise from the superposition of numerous unresolved astrophysical sources~\cite{Farmer:2003pa,Taylor:2012db}. Detecting the SGWB is of fundamental importance for understanding the very early history of the universe, fundamental physical laws, and the evolution of massive black holes.

One of the most effective methods for detecting the SGWB is to use multiple detectors for cross-correlation analysis~\cite{Allen:1997ad,Thrane:2013oya}. This method correlates the data streams from different detectors, effectively suppressing correlated instrumental noise and thereby extracting weak, common gravitational-wave signals. Applying the cross-correlation method to data from two OLC detectors can significantly enhance the sensitivity for detecting the SGWB~\cite{Wang:2024tnk}. The sensitivity of a detector network depends on the detector noise as well as the cross-correlation response. The cross-correlation response is fundamentally determined by the geometric factors of a detector pair, including their relative distance and orientation. The cross-correlation response function of a detector network, sometimes referred to as the overlap reduction function (ORF), is used to quantify these geometric factors. 
Currently, analytical expressions for precisely calculating the ORF have been derived in previous work~\cite{Hu:2025fev,Hu:2022ujx}. However, research on how to find the optimal geometric configuration for an OLC detector network that yields high response and effectively suppresses local noise remains an area of active investigation and presents significant challenges~\cite{Wang:2024tnk}.

In this paper, we explore the relationship between the ORF of an OLC detector network and its configuration geometry. This is achieved by identifying configuration transformations that leave the module of ORF invariant, and by numerically computing and comparing the ORF under different geometric characteristic parameters. The structure of this paper is as follows: In Sec.~\ref{Sec2}, we review the fundamental theory of the overlap reduction functions of OLC detector network. In Sec.~\ref{Sec3}, we explore ORF-invariant configuration transformations of the OLC detector network. In Sec.~\ref{Sec4}, we numerically compare the ORFs of an isosceles trapezoid configuration of the OLC detector network at different separations and different included angles. In Sec.~\ref{Sec5}, based on a practically feasible orbital configuration, we calculate the strain spectral sensitivity and noise energy density spectrum of the OLC detector network under that configuration, and compare them with those of space-based laser interferometric detectors such as LISA, Taiji, and TianQin. In Sec.~\ref{Sec6}, we summarize the work and discuss the results.

\section{ORF of OLC detector network}
\label{Sec2}

\begin{figure}
    \includegraphics[width=7cm]{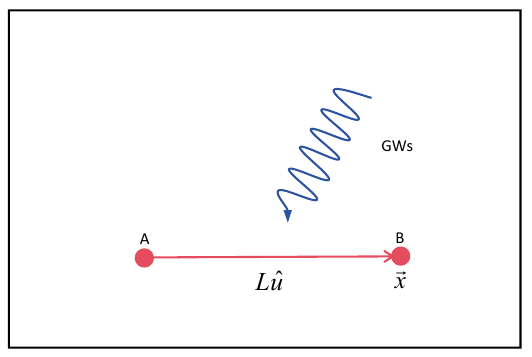}
\caption{Single-arm OLC detector configuration. An OLC detector consists of a one-way laser link and two satellites equipped with optical lattice atomic clocks.}
    \label{fig:singleOLC}
\end{figure}

In Ref.~\cite{Kolkowitz:2016wyg}, the single-arm OLC gravitational wave detector shown in Fig.~\ref{fig:singleOLC} is proposed,  which consists of two drag-free satellites separated by a distance $L$ and connected by a single laser link. Each satellite carries an optical lattice atomic clock. The time-frequency signal from the optical lattice atomic clock on satellite A is transmitted to satellite B via the inter-satellite laser link and compared with that from the optical lattice atomic clock on satellite B. When a gravitational wave passes through, it induces a Doppler shift in the time-frequency signal transmitted from A to B, thereby altering the clock difference observed in the time-frequency comparison signal at satellite B.

The gravitational wave background generates weak stochastic signals in an OLC detector $i$. To linear order, the SGWB signal recorded by the detector reads:
\begin{equation}
    s_i(t) = \sum_{A=+,\times} \int_{-\infty}^{\infty} df \int d^2\hat{n}\,R_i^A(f,\hat{n})\,
    \tilde{h}_A(f,\hat{n})\,e^{\text{i} 2\pi  f t}
     \,,
\end{equation}
where $R^A(f,\hat{n})$ is the response function of the detector to gravitational waves in the A polarization mode, $\tilde{h}_A(f,\hat{n})$ is the amplitude in the frequency domain of a gravitational wave propagating in the direction $\hat{n}$.
One single-arm OLC detector $i$ can be regarded as a one-way link detector, so its response function is:
\begin{equation}\label{R_OLC}
    R_{i}^A(f,\hat{n}) = \frac{1}{2}\,\hat{u}_i^a \hat{u}_i^b\, {\bf{e}}_{ab}^A \,\mathcal{T}_{i}(\hat{u}_i \cdot \hat{n}, f)\,e^{-\text{i} 2\pi  f \hat{n}\cdot\vec{x}_i/c},
\end{equation}
where $\vec{x}$ is the position of the spacecraft measuring the signal, for example, the position of spacecraft B in Fig.~\ref{fig:singleOLC}. $\mathcal{T}_{i}(\hat{u} \cdot \hat{n}, f)$ is the transfer function of a one-way link. In frequency shift measurements, its expression is~\cite{Hu:2025fev}:
\begin{equation}\label{trans}
    \mathcal{T}_{i}(\hat{u}_i \cdot \hat{n}, f) =  \frac{\text{i}2\pi f L_i}{c} \mathrm{sinc} \Big[ \frac{\pi f L_i}{c} (1-\hat{u}_i \cdot \hat{n}) \Big] e^{-\text{i}\frac{\pi f L_i}{c} (1-\hat{u}_i 
\cdot \hat{n})},
\end{equation}
where $L$ is the arm length of the detector, and $\hat{u}$ is the unit vector in the direction of laser propagation.

For an isotropic, unpolarized, stationary Gaussian gravitational wave background, its auto-correlation is:
\begin{equation}\label{eq2-39}
     \langle
    \tilde{h}_A(f,\widehat{\Omega})
    \tilde{h}_{A'}^*(f',\widehat{\Omega}')\rangle
    =\frac{1}{2}\delta(f-f')
    \frac{\delta^2(\widehat{\Omega},\widehat{\Omega}')}
    {4\pi}\delta_{AA'} \, S_h(f).
\end{equation}
Here $S_h(f)$ is the one-sided power spectral density of the gravitational wave backgound, satisfying $S_h(f)=S_h(-f)$.

\begin{figure}
    \includegraphics[width=7cm]{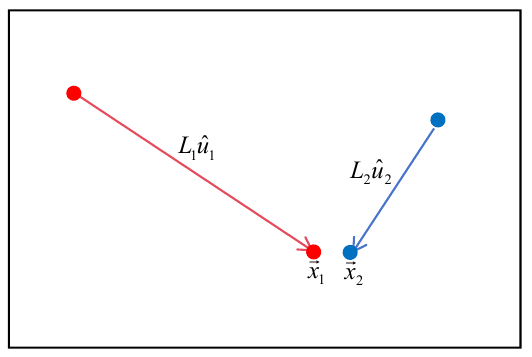}
\caption{The cross-correlation configuration between OLC detector 1 (red) and OLC detector 2 (blue).}
    \label{OLCpair}
\end{figure}

Two sets of OLC detectors can be used to improve the sensitivity to the SGWB via cross-correlation methods~\cite{Wang:2024tnk}, as shown in Fig.~\ref{OLCpair}. The cross-correlation between the signals of OLC detectors 1 and 2 is given by the following formula:
\begin{equation}\label{h1h2}
    \langle {s}_1(t) {s}_{2}(t') \rangle=\frac{1}{2}\int_{-\infty}^{\infty} df \,\Gamma_{12}(f) S_h(f) e^{\text{i} 2\pi  f (t-t')},
\end{equation}
where $\Gamma_{12}(f)$ is the ORF of OLC detectors 1 and 2. More generally, $\Gamma_{ij}(f)$ can be defined as
\begin{equation}
    \Gamma_{ij}(f) =\sum_{A=+,\times} \int \frac{d^2\hat{n}}{4\pi} R_i^{A}(f,\hat{n}) R_j^{A*}(f,\hat{n}),
\end{equation}
where $R_i^{A}(f,\hat{n})$ is the response function of the detector $i$ in Eq.~\eqref{R_OLC}.
The ORF can be interpreted as the weighted average of the responses of the two detectors to gravitational waves incident from all directions across the sky.
According to Eq.~\eqref{h1h2}, the upper bound of $\langle {s}_1(t) {s}_{2}(t') \rangle$ is proportional to the absolute value $|{\Gamma}_{12}(f)|$. For a given power spectrum $S_h(f)$, to maximize $|\langle {s}_1(t) {s}_{2}(t') \rangle|$, $|{\Gamma}_{12}(f)|$ must be as large as possible.
In fact, $|{\Gamma}_{12}(f)|$ quantifies the coherence strength of the responses of the two detectors to isotropic stochastic gravitational waves.

The geometric configuration of a single OLC detector can be described by three parameters: the endpoint position of the laser link $\vec{x}_{i}$, the arm length $L_i$, and the direction vector $\hat{u}_{i}$, collectively denoted as $\{\vec{x}_{i},L_i\hat{u}_{i}\}$. Similarly, the configuration of two OLC detectors is defined by $\{\vec{x}_{1},L_1\hat{u}_{1},\vec{x}_{2},L_2\hat{u}_{2}\}$. Fig.~\ref{OLCpair} schematically illustrates the configuration for calculating the cross-correlation response of an OLC detector pair.

\section{Equivalent configurations}
\label{Sec3}
 
Given any two OLC cross-correlation configurations $\{\vec{x}_{1},\allowbreak L_1 \hat{u}_{1}, \vec{x}_{2}, L_2\hat{u}_{2}\}$ and $\{\vec{x}_{3},L_3\hat{u}_{3},\vec{x}_{4},L_4\hat{u}_{4}\}$, through numerical integration, we can obtain the function curves of $|{\Gamma}_{12}(f)|$ and $|{\Gamma}_{34}(f)|$
for these two detector networks, and we can compare their ORF modulus at a given frequency $f$. However, since all configurations form a configuration space, it is impossible to exhaustively compare every configuration one by one.
Moreover, it is difficult to directly judge the modulus of the ORF from the analytical expression before ORF integration. We therefore adopt a different strategy: instead of directly comparing the modulus of two ORFs, we explore the relationship between configurations with the same ORF modulus, i.e., we look for configuration transformations that leave the ORF modulus invariant.

For a given OLC detector network configuration $\{\vec{x}_{1},\allowbreak L_1 \hat{u}_{1}, \vec{x}_{2}, L_2\hat{u}_{2}\}$, we aim to identify another OLC detector network configuration $\{\vec{x}_{3},L_3\hat{u}_{3},\vec{x}_{4},L_4\hat{u}_{4}\}$ such that their ORF satisfies
\begin{align} \label{equal}
|{\Gamma}_{12}(f)| = |{\Gamma}_{34}(f)|. 
\end{align}

We can easily identify three types of solutions:
The first type of solution, $\{\vec{x}_{2},L_2\hat{u}_{2},\vec{x}_{1}, L_1\hat{u}_{1}\}$, is generated from the original configuration $\{\vec{x}_{1},L_1\hat{u}_{1},\vec{x}_{2},L_2\hat{u}_{2}\}$ by swapping the labels 1 and 2 of the two OLC detectors.
The second type of solution, $\{\vec{x}_{1}+\vec{a},\,L_1\hat{u}_{1},\,\vec{x}_{2}+\allowbreak \vec{a},\,L_2\hat{u}_{2}\}$, is generated by translating the original configuration $\{\vec{x}_{1},L_1\hat{u}_{1},\vec{x}_{2},L_2\hat{u}_{2}\}$ by arbitrary vector $\vec{a}$.
The third type of solution, $\{\widehat{O}\vec{x}_{1},L_1\widehat{O}\hat{u}_{1},\widehat{O}\vec{x}_{2},  \allowbreak L_2\widehat{O}\hat{u}_{2}\}$, is generated from the original configuration  $\{\vec{x}_{1},\allowbreak L_1\hat{u}_{1},\vec{x}_{2},L_2\hat{u}_{2}\}$ via an orthogonal transformation $\widehat{O}$.
The first type of solution inherently possesses this property in the definition of the ORF, while the second and third types of solutions are essentially changes of the coordinate system. While keeping the modulus of the ORF unchanged, these three types of solutions do not alter the original configuration; therefore, they are all trivial transformations that leave the ORF modulus invariant.


To identify solutions corresponding to nontrivial configurations, we reformulate the ORF of OLC detector pair in the following form:
\begin{align}
&{\Gamma}_{12}(f)
={\displaystyle \sum_{\lambda}} \int \frac{d^2 \hat{n}}{4 \pi}\hat{u}_1^a\,\hat{u}_1^b\,  {\epsilon}^{\lambda}_{ab}(\hat{n})\,\hat{u}_2^c\,\hat{u}_2^d\, {\epsilon}^{\lambda}_{cd}(\hat{n})\nonumber\\
&\times\mathrm{sinc} \Big[ \frac{\pi f L_1}{c} (1-\hat{u}_1\cdot\hat{n}) \Big] \mathrm{sinc} \Big[\frac{\pi f L_2}{c} (1-\hat{u}_2 \cdot \hat{n}) \Big]\nonumber\\
&\times  L_1 L_2  \left(\frac{\pi f }{c}\right)^2 e^{\text{i} \alpha_{12}}e^{\text{i} \beta_{12}}\label{eq5}
,
\end{align}
where $\alpha_{12}\equiv {2 \pi f} ( \frac{L_1}{2}\hat{u}_1- \frac{L_2}{2}\hat{u}_2-\vec{x}_1+\vec{x}_2 
)\cdot \hat{n}$/$c$, and $\beta_{12}\equiv\allowbreak \pi f(L_2-L_1)$/$c$.
To find the solutions satisfying Eq.~\eqref{equal}, our strategy here is to allow the integrand of ${\Gamma}_{12}(f)$ to differ from that of ${\Gamma}_{34}(f)$ via an overall $O(2)$ transformation in the complex plane. 
According to Eq.~\eqref{eq5}, we find that the factors preceding the exponential factor are real-valued. Hence, we first equate these real factors for ${\Gamma}_{12}(f)$ and ${\Gamma}_{34}(f)$, which readily yields two conditions:
\begin{align}
L_3=L_1,~\hat{u}_{3}=\hat{u}_{1},~ 
L_4=L_2,~\hat{u}_{4}=\hat{u}_{2}\label{eq6},
\\
L_3=L_2,~\hat{u}_{3}=\hat{u}_{2},~ 
L_4=L_1,~\hat{u}_{4}=\hat{u}_{1}.\label{eq7}
\end{align}
Substituting these two conditions into the exponential factors yields the corresponding solutions, respectively. For convenience, we denote the original configuration $\{\vec{x}_{1},\,L_1\hat{u}_{1},\,\vec{x}_{2},\,L_2\hat{u}_{2}\}$ as configuration \{A\}.

When Eq.~\eqref{eq6} is adopted as a condition,  we have $\beta_{12}=\beta_{34}$, and again two subcases arise. First, if $\alpha_{12}=\alpha_{34}$, then
\begin{align}
\vec{x}_{1}-\vec{x}_{2}=\vec{x}_{3}-\vec{x}_{4},
\end{align}
and the resulting configuration is
\begin{align}
\{\vec{x}_{1}+\vec{a},\,L_1\hat{u}_{1},\,\vec{x}_{2}+\vec{a},\,L_2\hat{u}_{2}\},    
\end{align}
which is identical to the original configuration \{A\} (a pure translation). Second, if $\alpha_{12}=-\alpha_{34}$, then
\begin{align}\label{eq8}
(\vec{x}_{2}-L_2\hat{u}_{2})-(\vec{x}_{1}-L_1\hat{u}_{1})=\vec{x}_{3}-\vec{x}_{4},
\end{align}
and the resulting configuration is
\begin{align}
\{\vec{x}_{2}-L_2\hat{u}_{2}+\vec{a},\,L_1\hat{u}_{1},\,\vec{x}_{1}-L_1\hat{u}_{1}+\vec{a},\,L_2\hat{u}_{2}\},
\end{align}
which we label as configuration \{B\}. 

When Eq.~\eqref{eq7} is adopted as a condition, we have $\beta_{12}=-\beta_{34}$, two subcases arise. First, if $\alpha_{12}=-\alpha_{34}$, then 
\begin{align}
\vec{x}_{1}-\vec{x}_{2}=\vec{x}_{4}-\vec{x}_{3},
\end{align}
and the resulting configuration is
\begin{align}
    \{\vec{x}_{2}+\vec{a},\,L_2\hat{u}_{2},\,\vec{x}_{1}+\vec{a},\,L_1\hat{u}_{1}\}.
\end{align}
This is essentially the original configuration \{A\} with the labels 1 and 2 swapped. Second, if $\alpha_{12}=\alpha_{34}$, then
\begin{align}\label{eq10}
(\vec{x}_{2}-L_2\hat{u}_{2})-(\vec{x}_{1}-L_1\hat{u}_{1})=\vec{x}_{4}-\vec{x}_{3},
\end{align}
and the resulting configuration is
\begin{align}
\{\vec{x}_{1}-L_1\hat{u}_{1}+\vec{a},\,L_2\hat{u}_{2},\,\vec{x}_{2}-L_2\hat{u}_{2}+\vec{a},\,L_1\hat{u}_{1}\}.   
\end{align}
We label this configuration as configuration \{C\}, which can be obtained by exchanging the labels 3 and 4 in configuration \{B\}.
Thus configuration \{C\} can be identified with configuration \{B\}. 
\begin{figure}
    \centering
    \includegraphics[width=8cm]{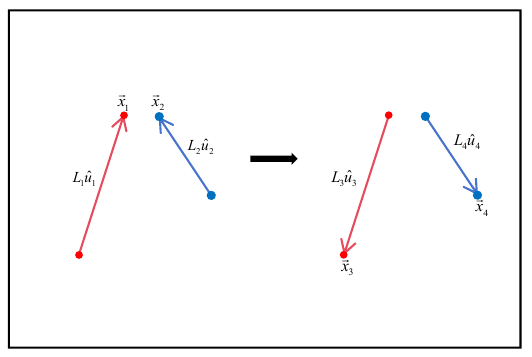}
    \caption{A nontrivial transformation that leaves the modulus of ORF invariant. The original configuration \{A\} (Detectors 1 \& 2) transforms into configuration \{B\} (Detectors 3 \& 4) equivalently. This nontrivial transformation can be summarized as swapping the emitting and receiving ends for both detector 1 and detector 2.}
    \label{nontrivial}
\end{figure}

As above discussion, starting from an arbitrary original configuration \{A\}, we finally obtain the configuration \{B\} with the modulus of the ORF invariant.
Without loss of generality, we may set an arbitrary vector $\vec{a}=L_1 \hat{u}_1+L_2 \hat{u}_2-\vec{x}_1-\vec{x}_2$.  Applying an orthogonal transformation diag$\{-1,-1,-1\}$ (which does not change the ORF) to  configuration \{B\} yields:
\begin{equation}
\begin{aligned}
&\{\vec{x}_{3},\,L_3\hat{u}_{3},\,\vec{x}_{4},\,L_4\hat{u}_{4}\}\\
=&\,\{\vec{x}_1-L_1\hat{u}_{1},\,-L_1\hat{u}_{1},\,\vec{x}_{2}-L_2\hat{u}_{2},\,-L_2\hat{u}_{2}\}.
\end{aligned}
\end{equation}

Here, $\vec{x}_3=\vec{x}_1-L_1\hat{u}_{1}$, meaning that the position where detector 3 receives the laser is the same as the position where detector 1 emits the laser;  and $L_3\hat{u}_{3}=-L_1\hat{u}_{1}$, meaning that the laser propagation direction of detector 3 is opposite to that of detector 1, with their arm lengths being equal. 
In this case, the signal of detector 3 is equivalent to that of detector 1 measured in the opposite direction.
Similarly, from $\vec{x}_4=\vec{x}_2-L_2\hat{u}_{2}$ and $L_4\hat{u}_{4}=-L_2\hat{u}_{2}$, it follows that the signal of detector 4 is equivalent to that of detector 2 measured in the opposite direction.

In Fig.~\ref{nontrivial}, we illustrate how the corresponding equivalent configuration \{B\} is constructed from an arbitrary original configuration \{A\}. In fact, this nontrivial transformation does not change the relative distance or relative angle between the two OLC detectors within the configuration, but merely swaps the emitting and receiving ends of each OLC detector.

Applying this nontrivial transformation again to configuration \{B\} recovers the initial configuration \{A\}, as shown by the following derivation:
\begin{equation}
\begin{aligned}
&\{\vec{x}_{5},L_5\hat{u}_{5},\vec{x}_{6},L_6\hat{u}_{6}\}\\
=&\{\vec{x}_3-L_3\hat{u}_{3},-L_3\hat{u}_{3},\vec{x}_{4}-L_4\hat{u}_{4},-L_4\hat{u}_{4}\}\\
=&\{\vec{x}_1-L_1\hat{u}_{1}-L_3\hat{u}_{3},-L_3\hat{u}_{3},\vec{x}_{2}-L_2\hat{u}_{2}-L_4\hat{u}_{4},-L_4\hat{u}_{4}\}\\
=&\{\vec{x}_1-L_1\hat{u}_{1}+L_1\hat{u}_{1},L_1\hat{u}_{1},\vec{x}_{2}-L_2\hat{u}_{2}+L_2\hat{u}_{2},L_2\hat{u}_{2}\}\\
=&\{\vec{x}_{1},L_1\hat{u}_{1},\vec{x}_{2},L_2\hat{u}_{2}\}.
\end{aligned}
\end{equation}

In general, when $L_1 \neq L_2$ and $L_3 \neq L_4$, i.e., the two detectors in a detector network have unequal arm lengths, the modulus of the two ORFs before and after the nontrivial transformation are equal, but their arguments are not equal; they differ by a value related to the arm length difference, which is reflected in the exponential factors $\beta_{12}$ and $\beta_{34}$. In particular, when $L_1 = L_2$ and $L_3 = L_4$, i.e., the two detectors in a detector network have equal arm lengths, we have $\beta_{12} = \beta_{34} = 0$, and the nontrivial transformation satisfies either $\alpha_{12} = \alpha_{34}$ or $\alpha_{12} = -\alpha_{34}$. Correspondingly, either $\Gamma_{12}(f) = \Gamma_{34}(f)$ or $\Gamma_{12}(f) = \Gamma_{34}^*(f)$. In this case, the two ORFs before and after the nontrivial transformation are either equal or complex conjugates of each other.

\section{ORFs for different configurations}
\label{Sec4}
 
In this section, we compare the ORFs of an isosceles trapezoid configuration of the OLC
detector network at different separations and different included angles.
Firstly, we define
\begin{equation}
    \vec{s} = \hat{u}_1+\hat{u}_2\,,
\end{equation}
and
\begin{equation}
    \vec{m}_{12}=(\vec{x}_2 -\frac{L}{2}\hat{u}_2)-(\vec{x}_1-\frac{L}{2}\hat{u}_1)\,,
\end{equation}
where $\vec{m}_{12}$ appears in the factor $\alpha_{12}$ and can be viewed as the vector pointing from the center of mass of detector 1 to that of detector 2. 

It can be shown that when $L_1=L_2$ and $\vec{s}\cdot\vec{m}_{12}=0$, the imaginary part of $\Gamma_{12}(f)$ is zero. (The derivation is given in Appendix \ref{AppA}). Denote the starting positions of the laser links of detector 1 and detector 2 as $\vec{y}_1$ and $\vec{y}_2$, which satisfy:
\begin{align}
    \vec{y}_1=\vec{x}_1-L\hat{u}_1\,,\\
    \vec{y}_2=\vec{x}_2-L\hat{u}_2\,.
\end{align}
Besides,
\begin{equation}
    |\vec{x}_2-\vec{y}_1|^2
    =\,|\vec{m}_{12}+\frac{L}{2}\vec{s}\,|^2=\,|\vec{m}_{12}|^2+L\vec{s}\cdot\vec{m}_{12}+\frac{L^2}{4}|\vec{s}|^2\,,
\end{equation}
\begin{equation}
    |\vec{y}_2-\vec{x}_1|^2
    =\,|\vec{m}_{12}-\frac{L}{2}\vec{s}\,|^2=\,|\vec{m}_{12}|^2-L\vec{s}\cdot\vec{m}_{12}+\frac{L^2}{4}|\vec{s}|^2\,,
\end{equation}
so
\begin{equation}
    |\vec{x}_2-\vec{y}_1|^2-|\vec{y}_2-\vec{x}_1|^2=2L\vec{s}\cdot\vec{m}_{12}=0.
\end{equation}

\begin{figure}
    \centering
    \includegraphics[width=7cm]{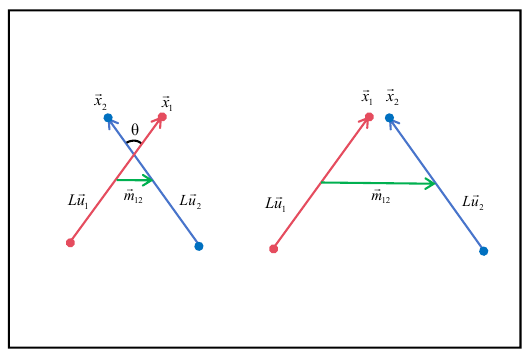}
    \caption{Schematic diagram of the isosceles trapezoid configurations. $\theta$ is the included angle, and $|\vec{m}_{12}|$ is the distance between the midpoints of the two laser links.}
    \label{fig:x}
\end{figure}

Therefore, $\vec{s}\cdot\vec{m}_{12}=0$ is equivalent to $|\vec{x}_2-\vec{y}_1|=|\vec{y}_2-\vec{x}_1|$. This implies that if the distance from the start point of the laser link of OLC detector 1 to the endpoint of the laser link of OLC detector 2 equals the distance from the start point of the laser link of OLC detector 2 to the endpoint of the laser link of OLC detector 1, and $L_1=L_2$, then their ORF is real. In particular, when the two OLC detectors lie in the same plane, the four spacecraft form an isosceles trapezoid (or an isosceles triangle), as shown in Fig.~\ref{fig:x}.

\begin{figure}
    \centering
    \includegraphics[width=7.5cm]{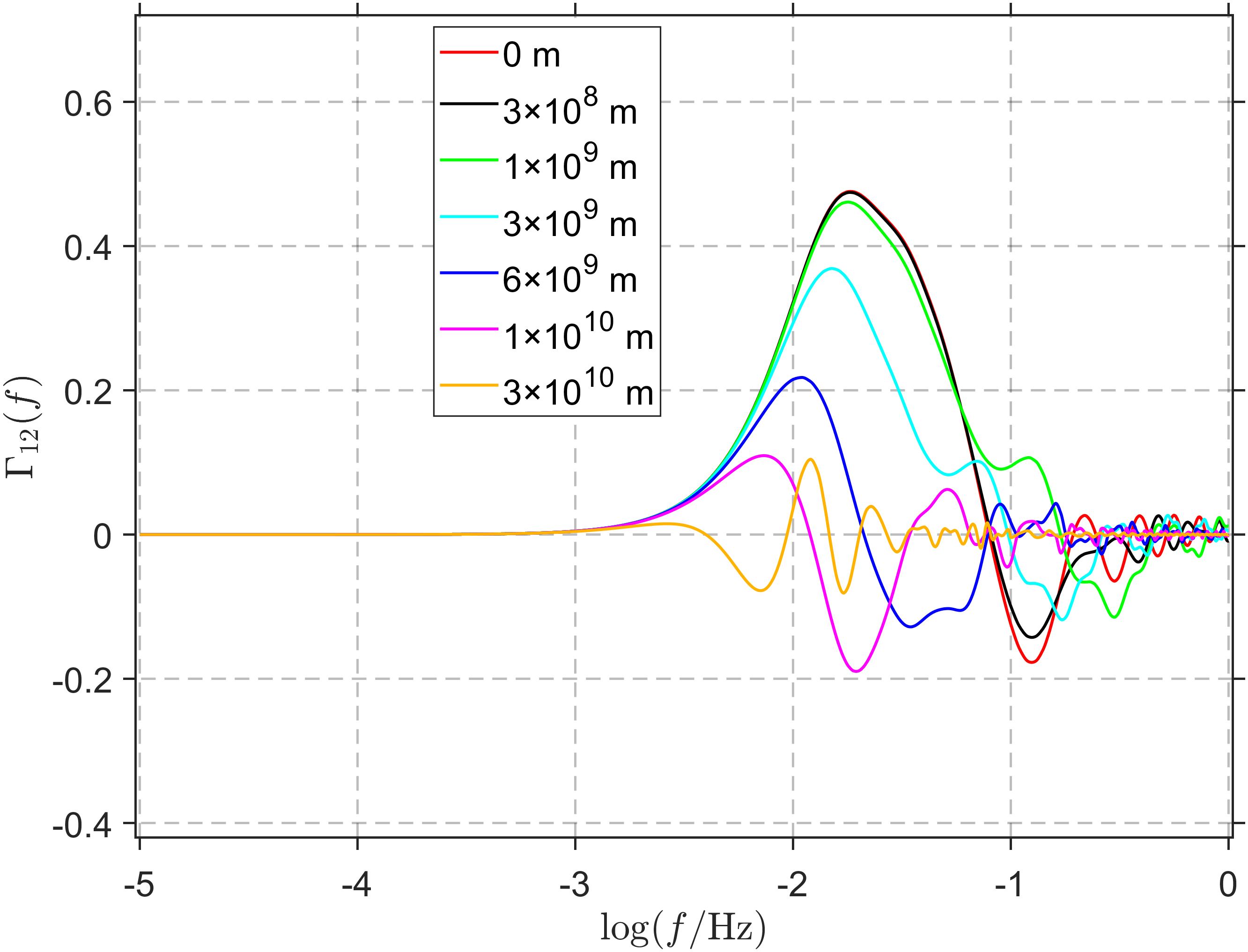}
    \caption{
    Comparison of ORFs for an isosceles trapezoid configuration in Fig.~\ref{fig:x}, at different distances $|\vec{m}_{12}|$, with an included angle of $\theta=20^{\circ}$ and an arm length of $L=10^{7}$ km}
    \label{fig:dfm12}
\end{figure}

In Fig.~\ref{fig:dfm12}, the ORFs of an isosceles trapezoid configuration with an included angle of 20$^{\circ}$ and an arm length of $10^{7}$ km are compared at different distances $|\vec{m}_{12}|$, where the ORF curves for separations of 0 and $3\times10^{5}$ km almost coincide in the low-frequency range below $0.1$Hz. In Fig.~\ref{fig:dfm12}, all ORF curves approach zero in the low-frequency range below $10^{-3}$ Hz, then increase with frequency and reach the first peak around $10^{-3}$ Hz. 
In the subsequent higher-frequency range, these ORFs oscillate as the frequency increases.
As $|\vec{m}_{12}|$ increases, the ORF reaches its peak at a lower frequency, and the peak value of the $|\Gamma_{12}(f)|$ also decreases.

\begin{figure*}[!htbp]
    \centering
    \begin{subfigure} 
        \centering
        \includegraphics[width=0.4\linewidth]{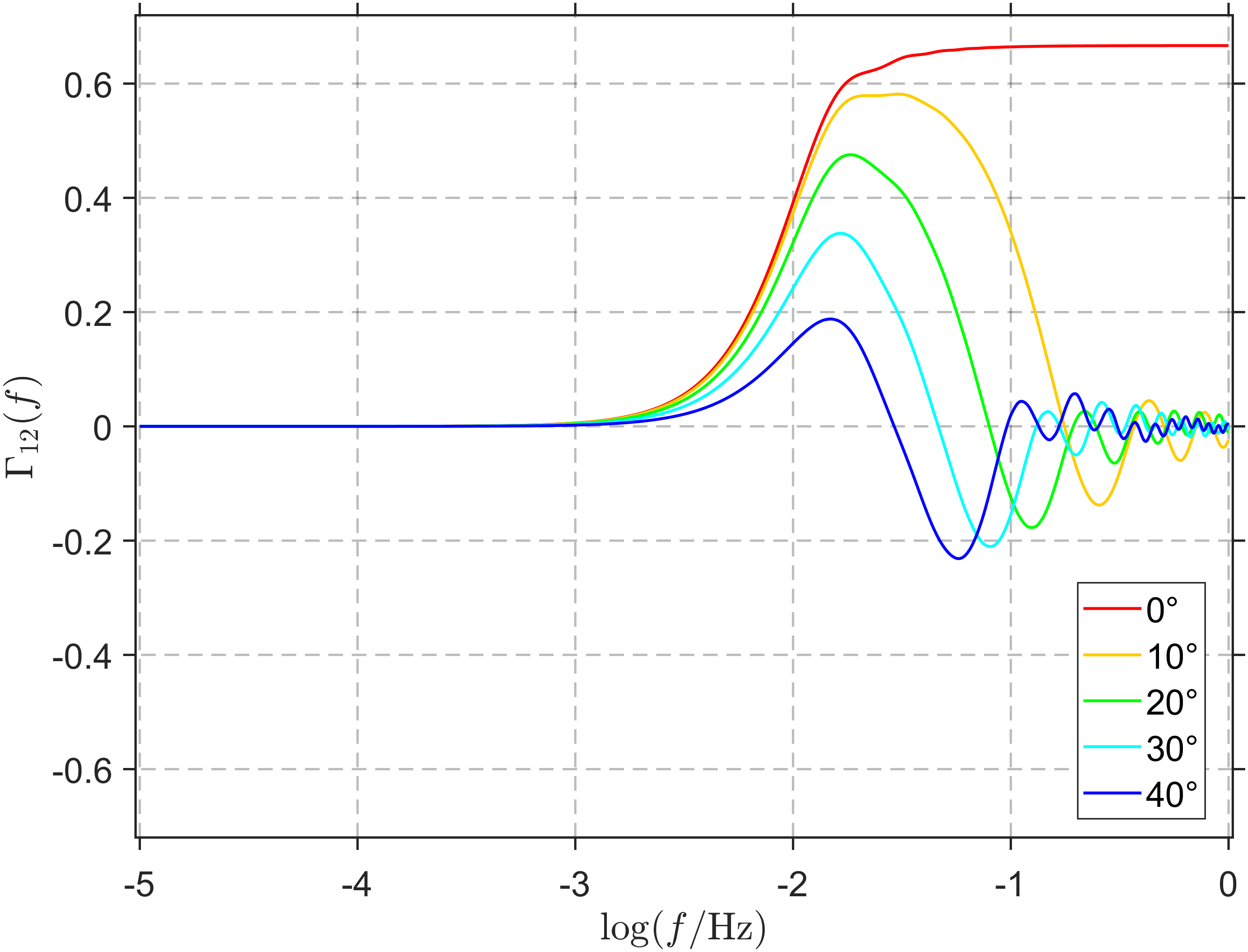}
    \end{subfigure}
    \hspace{-0.01\linewidth}
    \begin{subfigure} 
        \centering
        \includegraphics[width=0.4\linewidth]{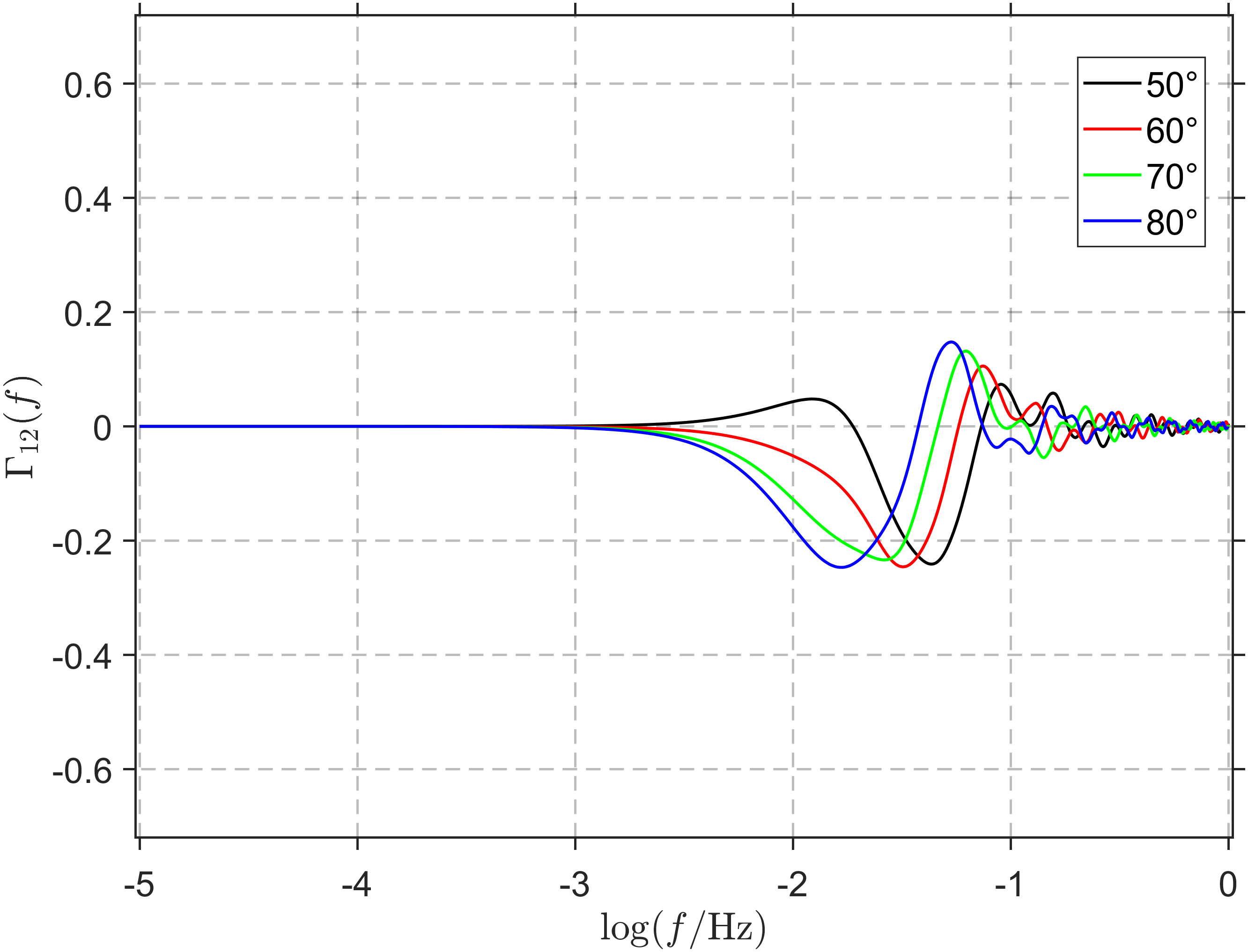}
    \end{subfigure}
    \begin{subfigure} 
        \centering
        \includegraphics[width=0.4\linewidth]{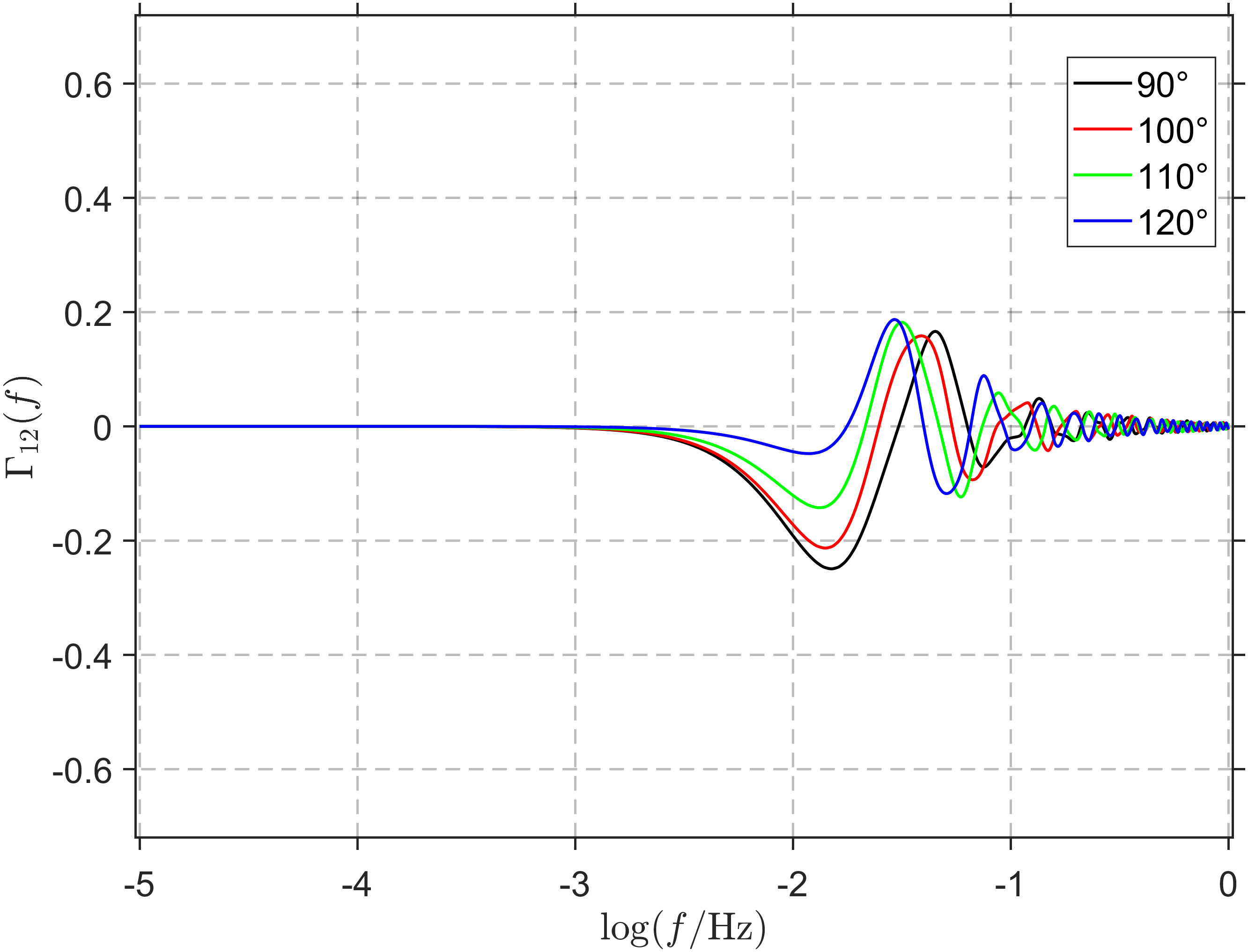}
    \end{subfigure}
    \hspace{-0.01\linewidth}
    \begin{subfigure} 
        \centering
        \includegraphics[width=0.4\linewidth]{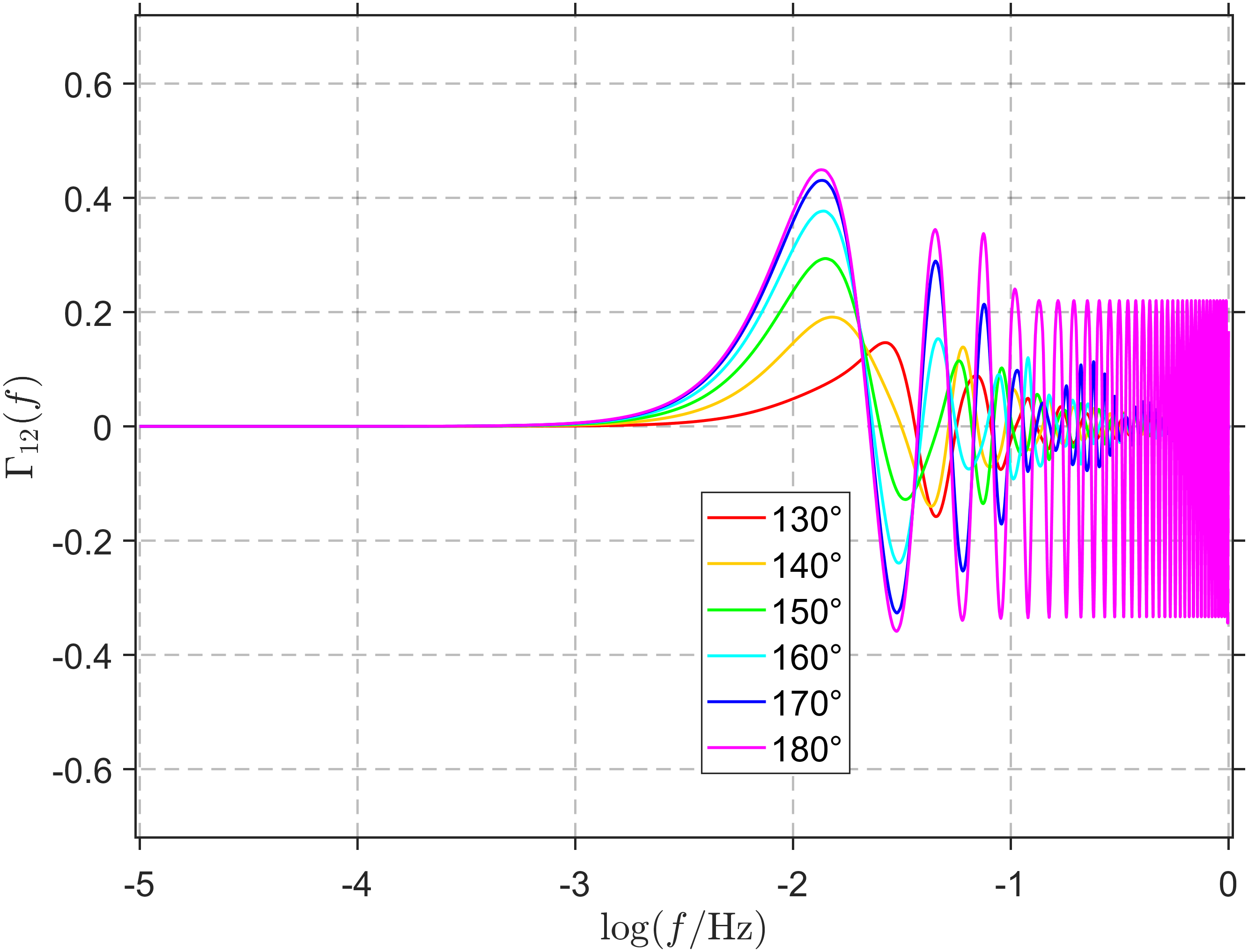}
    \end{subfigure}
    \caption{Comparison of ORFs for an isosceles trapezoid configuration in Fig.~\ref{fig:x} with $|\vec{m}_{12}|=0$ and arm length $L=10^{7}$ km at different included angles $\theta$.}
    \label{fig:x-0-180}
\end{figure*}

In Fig.~\ref{fig:x-0-180}, we compare the ORFs of an isosceles trapezoid configuration with $|\vec{m}_{12}|=0$ and an arm length of $10^{7}$ km for different included angles $\theta$. We focus on the first peak of the ORF from low frequencies for each angle: From 
0$^{\circ}$ to 50$^{\circ}$, the first peak decreases monotonically with increasing angle; From 60$^{\circ}$ to 80$^{\circ}$, the first peaks are all negative. From 90$^{\circ}$ to 120$^{\circ}$, the first peaks are also negative, but they increase monotonically with increasing angle. From 130$^{\circ}$ to 180$^{\circ}$, the first peaks are all positive and increase monotonically with increasing angle. The above trend is generally consistent with the angular dependence of the Hellings-Downs curve~\cite{Hellings:1983fr}:
$\Gamma_{\rm HD}(\gamma)=\frac{1}{2}+3\sin^2{\frac{\gamma}{2}} \ln{\sin{\frac{\gamma}{2}}}-\frac{1}{4}\sin^2{\frac{\gamma}{2}}$,
where $\gamma$ is the angle between the two directions.

\section{Sensitivity curves}
\label{Sec5}

In this section, we will calculate the sensitivity of an OLC detector network under a certain configuration and compare it with the sensitivities of laser interferometric gravitational wave detectors such as LISA, Taiji, and TianQin. In the following, we focus on the case where the two detectors have equal arm lengths, i.e., $L_1=L_2=L$.

The strain spectral sensitivity $\widetilde{h}_{n}(f)$ of a gravitational wave detector is defined as the square root of the effective noise power spectral density $S_n(f)$~\cite{Cornish:2001qi}:
\begin{equation}
\widetilde{h}_{n}(f)=\sqrt{S_n(f)}=\sqrt{\frac{N_m(f)}{{\cal R}_m(f)}}.\label{strain2}
\end{equation}
Here, ${\cal R}_m(f)$ is the sky averaged response function~\cite{Cornish:2001qi}
\begin{equation}
\mathcal{R}_{m}(f) = {\displaystyle \sum_{A=+,\times}} \int \frac{d^2 \hat{n}}{4 \pi} F_m^{\lambda}(\hat{n}, f) F_m^{\lambda}{}^*(\hat{n}, f),
\end{equation}
where $F_m^{\lambda}(\hat{n}, f)$ is the pattern function,
\begin{equation}
F_m^{\lambda}(\hat{n}, f) = D^{ab}_m(\hat{n}, f) {\epsilon}^{\lambda}_{ab}(\hat{n}).
\end{equation}
$D^{ab}_m(\hat{n}, f)$ is the detector response tensor,
\begin{equation}
D^{ab}_m(\hat{n}, f) = \frac{1}{2}\big[u^au^b  \mathcal{T}_m(\hat{u}\!\cdot\!\hat{n}, f)- v^a v^b  \mathcal{T}_m(\hat{v}\!\cdot\!\hat{n}, f)\big],
\end{equation}
and the transfer function is
\begin{equation}
\begin{aligned}
\mathcal{T}_m(\hat{u}\!\cdot\!\hat{n}, f) &=\frac{1}{2}\mathrm{sinc}\big[\frac{\pi f L}{c}(1-\hat{u}\!\cdot\!\hat{n})\big]e^{-\text{i}\frac{\pi f L}{c}(3+\hat{u}\cdot\hat{n})}\\
&+ \frac{1}{2}\mathrm{sinc}\big[\frac{\pi f L}{c}(1+\hat{u}\!\cdot\!\hat{n})\big]e^{-\text{i}\frac{\pi f L}{c}(1+\hat{u}\cdot\hat{n})}.
\end{aligned}    
\end{equation}

For LISA, Taiji, and TianQin, we choose the approximated Michelson-type channel noise model $N_m(f)$ as in Ref.\cite{Robson:2018ifk}
\begin{equation}
\begin{aligned}
{N}_{m}(f)=&\frac{A^2_{\text{OMS}}}{L^2}\left[1+\left(\frac{2\mathrm{mHz}}{f}\right)^4\right]
+\left[1+\mathrm{cos}^2\left(\frac{2\pi f L}{c}\right)\right]\\
\times &\frac{2 A^2_{\text{acc}}}{(2\pi f)^4 L^2}\left[1+\left(\frac{0.4 \mathrm{mHz}}{f}\right)^2\right] \left[1+\left(\frac{f}{8 \mathrm{mHz}}\right)^4\right].
\end{aligned}    
\end{equation}
The noise parameters are summarized in Table~\ref{t1}.
\begin{table}[!htbp]
    \centering
    \begin{tabular}{|c|c|c|c|}
        \hline
Parameters & ~LISA~ & ~Taiji~ & ~TianQin~\\
        \hline
$L$ (Gm) & 2.5  & 3  & 0.17    \\
        \hline
$A_{\text{OMS}}(\mathrm{pm/Hz^{1/2}})$ & 15  & 8  & 1  \\
        \hline
$A_{\text{acc}}(\mathrm{fm/s^2/Hz^{1/2}})$ & 3  & 3   & 1   \\
        \hline
    \end{tabular}
    \caption{The noise parameters of LISA~\cite{Robson:2018ifk},  Taiji~\cite{Luo:2021qji}, and TianQin~\cite{TianQin:2015yph} are summarized in this table.}
    \label{t1}
\end{table}

The strain spectral sensitivity of the OLC detector network can be written as~\cite{Wang:2024tnk}:
\begin{equation}\label{strain1}
\widetilde{h}_{12}(f) = \left[(\tau\Delta f)  \overline{\left(\frac{{\Gamma}_{12}(f)}{N_{\rm total}(f)}\right)^2}\right]^{-1/4},
\end{equation}
where $\overline{X(f)}$ denotes the average of $X(f)$ over a small interval centered at $f$ with width $\Delta f$, and $\tau$ is the observation time.
We adopt $\tau = 1$ year and frequency resolution $\Delta f = f/10$, as in Ref.~\cite{Wang:2024tnk}.

The total noise power spectral density for the OLC detector comprises three main contributions: quantum projection noise (QPN), photon shot noise (PSN), and acceleration noise (AN): $N_{\rm total}(f)=N_{\rm QPN}(f)+N_{\rm PSN}(f)+N_{\rm AN}(f)$, where
\begin{align}
 & N_{\rm QPN}(f)~=~\frac{2}{(2 \pi \nu)^2 N T } ~, \\
 & N_{\rm PSN}(f)~=~\frac{2  f^2 \Delta_L }{\nu^2 \big[ (2 \pi f)^2 +\eta P_L \Delta_L / (2\pi \hbar \nu) \big] } ~, \\
 & N_{\rm AN}(f)~=~\frac{S_L}{c^2 (2 \pi f)^2 } ~.
\end{align}
Here, we choose the parameter values as in Ref.~\cite{Kolkowitz:2016wyg},
\begin{align}
\label{parameters}
&
\nu = 430\,\text{THz}, ~N = 7 \times 10^6, ~T = 160\,\text{s},~ \eta=0.5,\nn\\
& \Delta_L = 30 {\mathrm{mHz}}, ~
P_L=3\,{\rm pW},  ~~
S_L = 9   \text{fm}^2 \, \text{s}^{-4} \, \text{Hz}^{-1}.
\end{align}

\begin{figure}[htbp]
    \centering
    \includegraphics[width=0.3\textwidth]{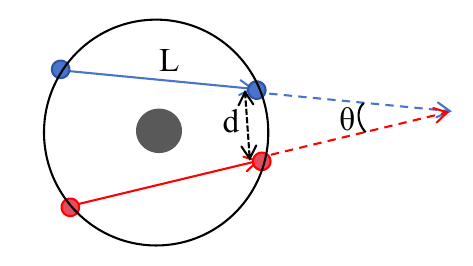}
    \caption{One orbital configuration of the OLC detector selected in this work. The four spacecraft of the OLC detector network are located on a common circle. The four-spacecraft formation forms an isosceles trapezoid. The angle between the baseline directions of the two detectors is $\theta$, their arm length is $L$, and the closest distance between the two detectors is $d$.}
    \label{orbit}
\end{figure}

In Ref.~\cite{Dhurandhar:2004rv}, it has been demonstrated that satellites located near the center of a circle with radius $R$ in a plane inclined at $60^\circ$ to the ecliptic can maintain relative stationarity, provided the orbital eccentricity remains small.
In order to ensure that the actual orbits of the four spacecraft in the OLC detector network all have small eccentricities, we can place the four spacecraft of the OLC detector network on a common circle. Here, we choose the orbital configuration shown in Fig.~\ref{orbit}. We choose the detector separation to be $d=5\times10^{3}$ km. This distance is adopted from Ref.~\cite{Wang:2024tnk} to reduce the correlation of local noise. To maximize the ORF of this configuration while maintaining reasonable source localization capability, we set $\theta$ $=25^\circ$.

\begin{figure}[!htbp]
    \centering
 \includegraphics[width=7cm]{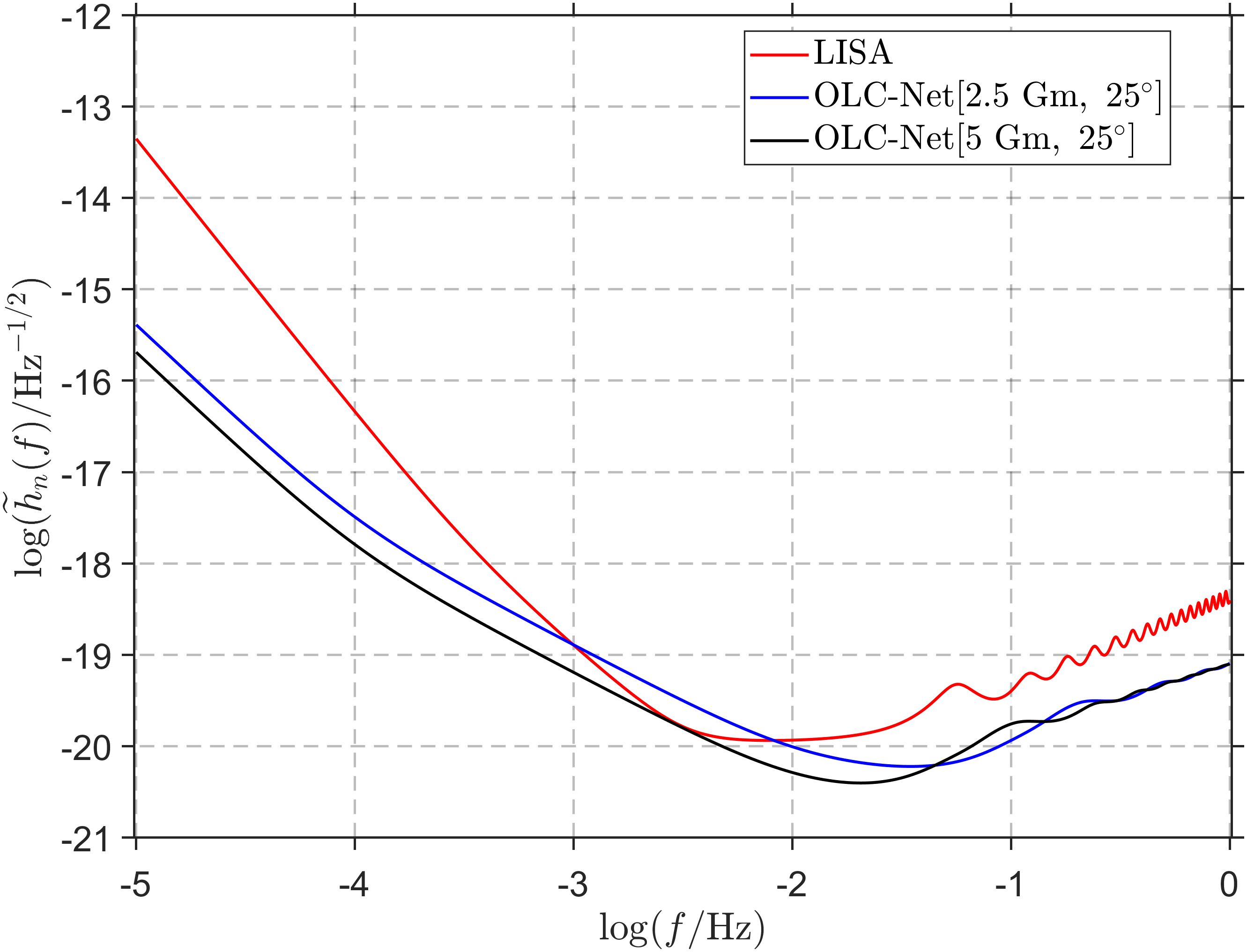}
 \includegraphics[width=7cm]{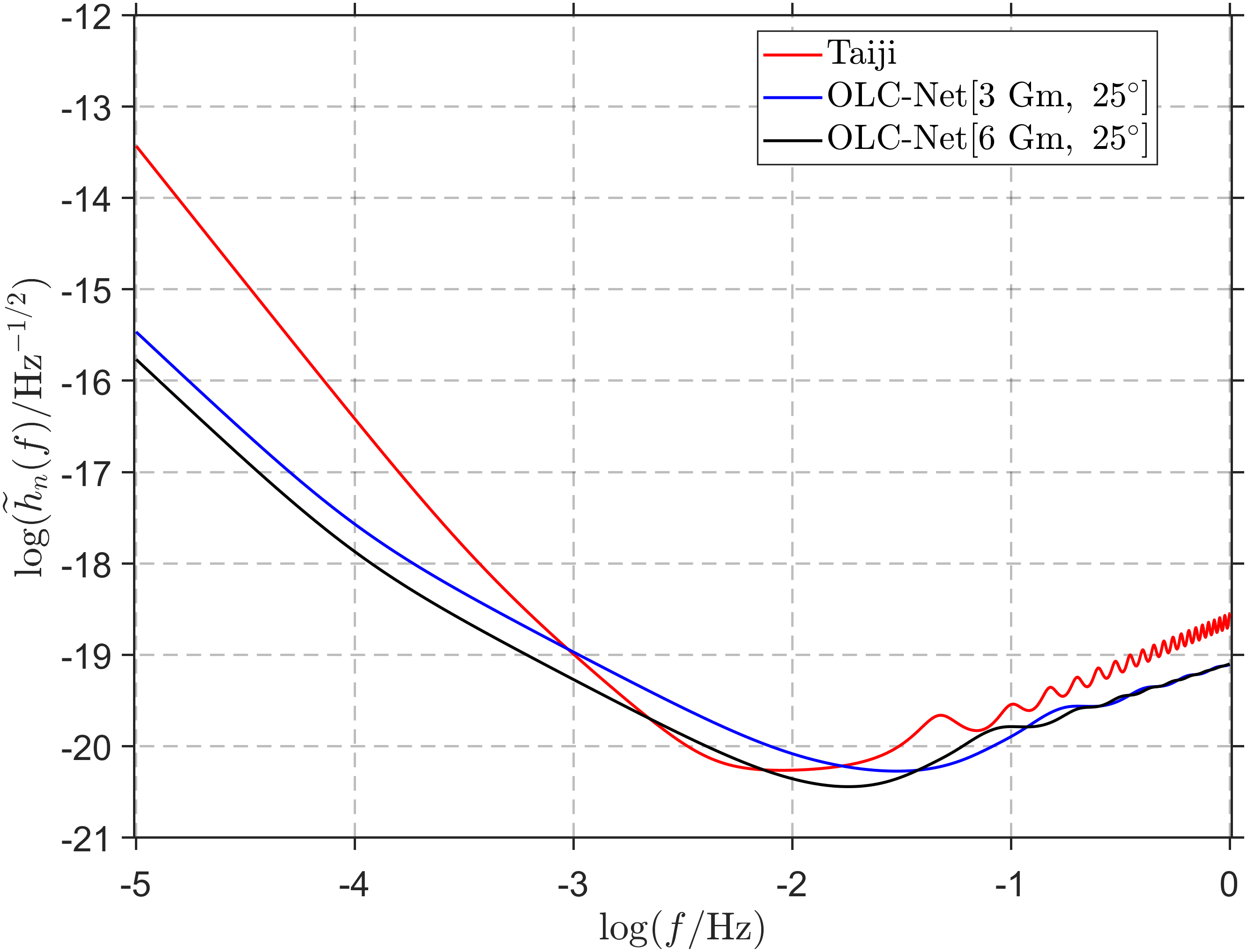}
 \includegraphics[width=7cm]{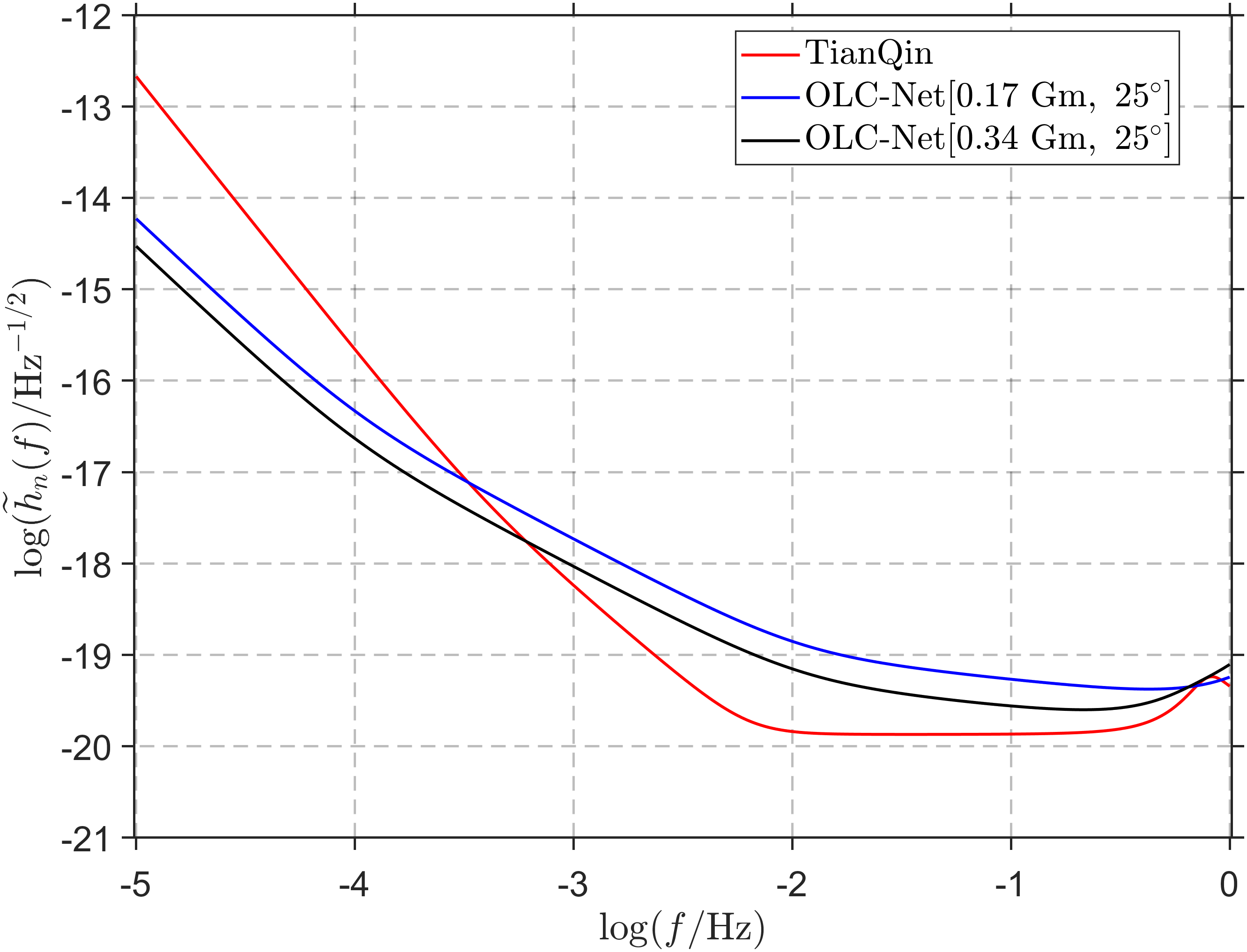}
\caption{Comparison of the strain spectral sensitivity in Eq.~\eqref{strain2} for LISA, Taiji, TianQin, and OLC detector network configuration in Fig.~\ref{orbit}. For the OLC network, we use Eq.~\eqref{strain1} and two arm lengths are considered: one that matches the triangular detector's arm, the other being twice as long.}
    \label{sensitivity}
\end{figure}


In Fig.~\ref{sensitivity}, we compare the strain spectral sensitivity of the OLC detector network configuration in Fig.~\ref{orbit} with that of the space-based laser interferometer gravitational wave detectors: LISA, Taiji, and TianQin. 
The strain spectral sensitivity $\widetilde{h}_{n}(f)$ in Eq.~\eqref{strain1} and  Eq.~\eqref{strain2}, can be further converted into the dimensionless noise energy density spectrum~\cite{Cornish:2001qi},
\begin{equation}
\Omega_nh^2=\frac{4\pi^2 f^3}{3(H_0/h)^2}\widetilde{h}_{n}^2(f).\label{eq21}
\end{equation}

\begin{figure}[htbp]
    \centering
\includegraphics[width=7cm]{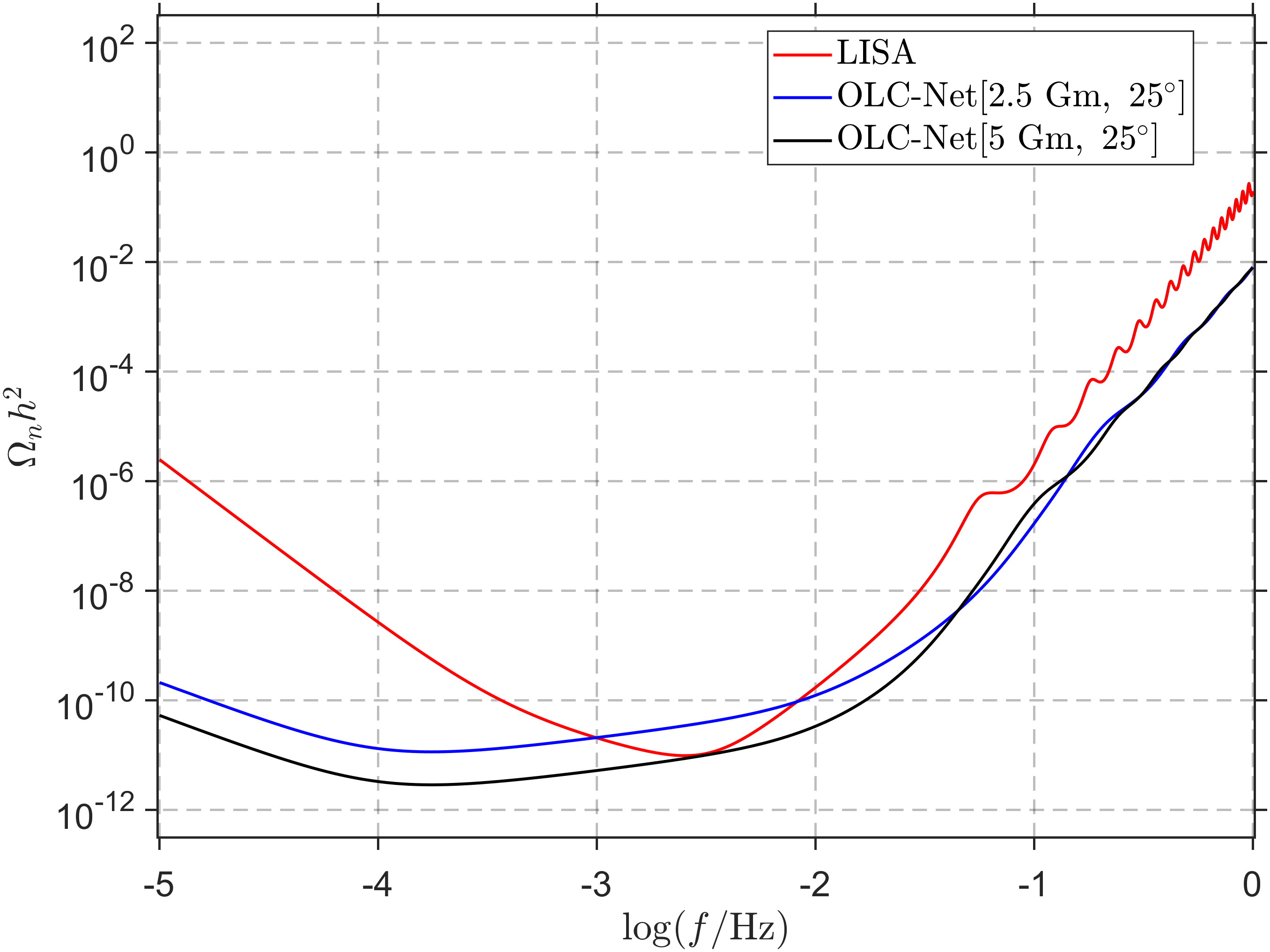}
\includegraphics[width=7cm]{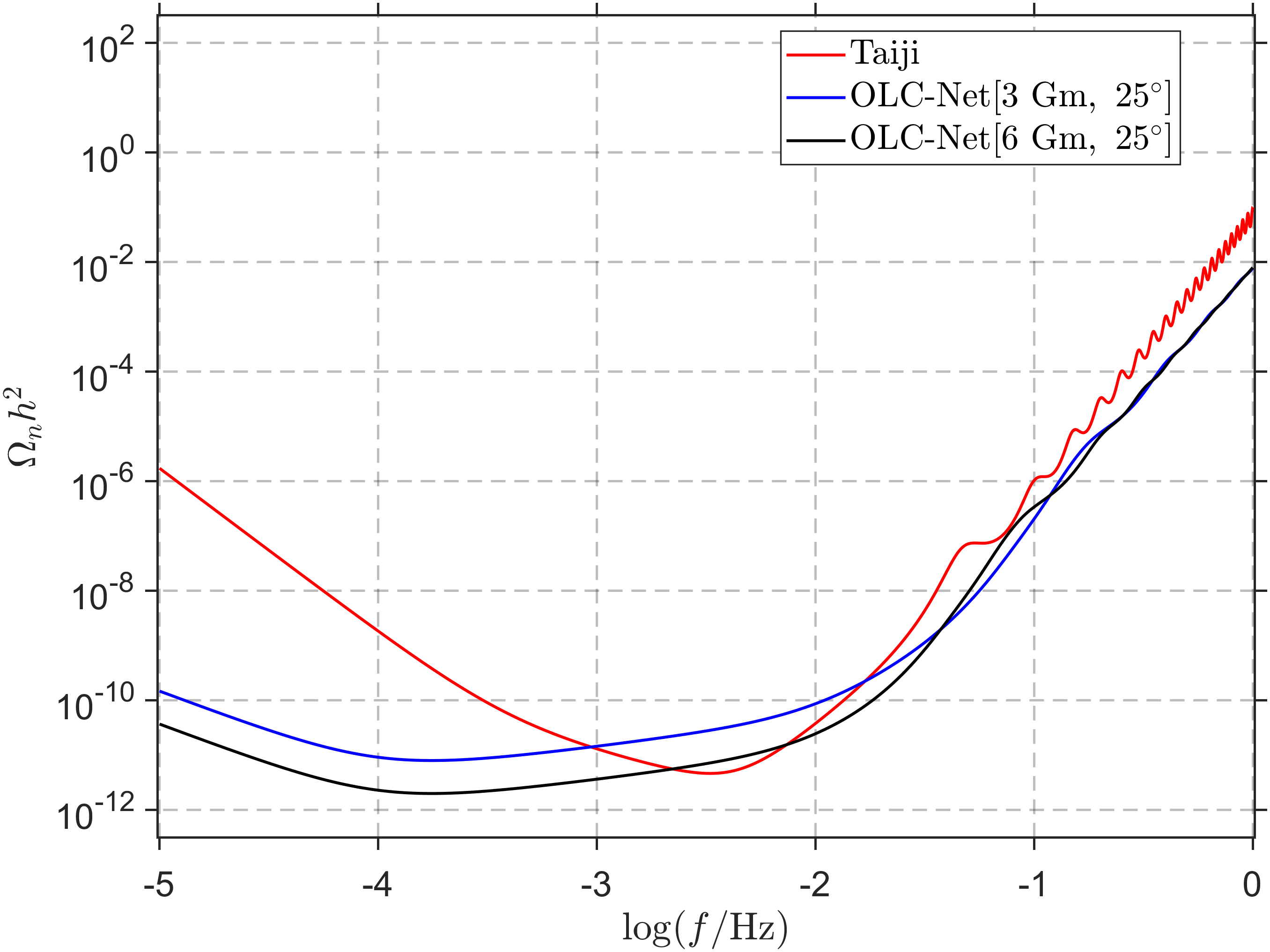}
\includegraphics[width=7cm]{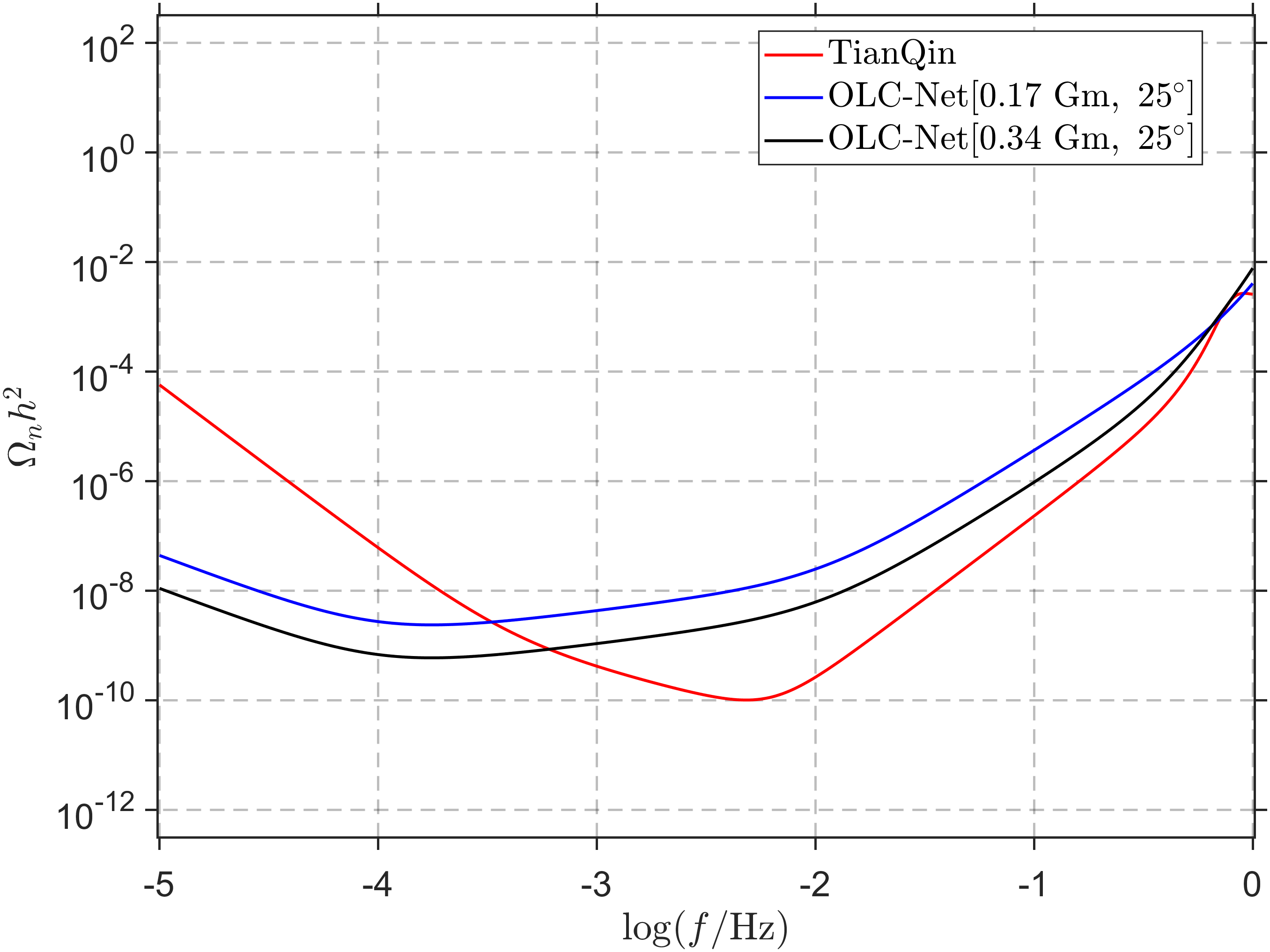}
\caption{Noise energy density curves of LISA, Taiji, TianQin and OLC detector network configuration in Fig.~\ref{orbit}.}
    \label{NoiseEnergy}
\end{figure}

Here, $H_0$ is the present-day Hubble parameter, conventionally written as $H_0 = 100\,h~\mathrm{km\,s^{-1}\,Mpc^{-1}}$.
To avoid dependence on the specific value of the dimensionless parameter $h$, we use $H_0/h\simeq 3.24\times 10^{-18}~\mathrm{s^{-1}}$, and the results of $\Omega_nh^2$ are plotted in Fig.~\ref{NoiseEnergy}. 
The OLC detector network configuration in Fig.~\ref{orbit} shows better sensitivity than LISA and Taiji in both the low and high frequency bands. The medium frequency regime around millihertz corresponds to the most sensitive band for LISA and Taiji, where they achieve relatively superior sensitivity. TianQin exhibits better strain spectral sensitivity than the OLC detector network configuration in Fig.~\ref{orbit} in the frequency range above $10^{-4}$ Hz, whereas the OLC configuration is superior in the low-frequency regime.

\section{Conclusion and discussion}
\label{Sec6}


This paper investigates the cross‑correlation response of an optical lattice clock (OLC) gravitational wave detector network to a stochastic gravitational wave background, with a focus on the relationship between the overlap reduction function (ORF) of the OLC detector network and its geometric configuration. The paper first reviews the basic theory of single-arm OLC detectors and OLC detector networks for SGWB detection. On the theoretical side, starting from the complex structure of the ORF integrand, this paper discusses under which configuration transformations the ORF modulus remains invariant.  
In addition to trivial transformations such as translation, rotation, and exchanging detector labels, we identify a nontrivial equivalent transformation (as shown in Fig.~\ref{nontrivial}): swapping the laser transmitting and receiving ends of each detector leaves the magnitude of the ORF unchanged.

The ORF integrand represents the inner product of the responses of two detectors to a gravitational wave from a given direction. In deriving this nontrivial configuration transformation, we set the integrands of $\Gamma_{12}(f)$ and $\Gamma_{34}(f)$ equal for the same propagation direction $\hat{n}$. Thus, this nontrivial transformation applies not only to the analysis of stochastic gravitational waves but also to the detection of plane gravitational waves from a single direction.  
Moreover, under the condition of equal arm lengths, the ORFs before and after the transformation not only have the same modulus but are either equal or complex conjugates of each other.  
Furthermore, we present a sufficient condition for the ORF to be real. The result shows that when the two OLC detectors have equal arm lengths and satisfy $\vec{s}\cdot\vec{m}_{12}=0$, the imaginary part of the ORF vanishes. This condition can also be expressed in a more intuitive geometric form: $|\vec{x}_2-\vec{y}_1|=|\vec{y}_2-\vec{x}_1|$ i.e., the distance from the starting point of the laser link of detector 1 to the endpoint of the laser link of detector 2 equals the distance from the starting point of the laser link of detector 2 to the endpoint of the laser link of detector 1.

In the numerical analysis part, we explore the variation of the ORF with detector separation and included angle for an isosceles trapezoid configuration. The results show that for a fixed included angle, as the separation between the centers of the two detectors increases, the ORF peak decreases and shifts toward lower frequencies. For a fixed separation satisfying 
$|\vec{m}_{12}|=0$, the variation of the ORF with the included angle is generally consistent with the Hellings-Downs curve.

Finally, based on a practically feasible orbital configuration, this paper calculates the strain spectral sensitivity and noise energy density spectrum of the OLC detector network under that configuration, and compares them with those of space‑based laser interferometric detectors such as LISA, Taiji, and TianQin. Compared with LISA and Taiji, the OLC network exhibits better sensitivity in both the low‑frequency and high‑frequency bands. Compared with TianQin, the OLC network has better sensitivity at low frequencies. Overall, this paper provides useful guidance for the design of OLC detector network configurations.

Based on the work presented in this paper, several challenges remain to be addressed in the future. For example, we can seek other nontrivial solutions and explore nontrivial transformations that leave the circular polarization components of the ORF $\Gamma_{ab}^{V}$ unchanged~\cite{Jiang:2025uga,Liu:2025hwn}.
In addition, when studying the variation of the ORF with the included angle and detector separation, we restrict the two detectors to lie in the same plane and impose $\vec{s}\cdot\vec{m}_{12}=0$.
In future work, more spatial configurations of OLC detectors can be explored, and their ORFs can be systematically compared and analyzed.
According to Ref.~\cite{Hu:2025fev}, the complex links in space-based laser interferometers such as LISA can be decomposed into combinations of one-way links. Therefore, in future work, we can attempt to apply the nontrivial transformation to these complex links and explore the properties of the complex links after the equivalent transformation.

\begin{acknowledgments}
This work is supported by the National Key Research and Development Program of China (No.~2023YFC2206200,~No.~2021YFC2201901), the National Natural Science Foundation of China (No.~12375059, No.~12575063),  and the Project of National Astronomical Observatories, Chinese Academy of Sciences (No.~E4TG6601). We thank Bo Wang for helpful discussions.
\end{acknowledgments}

\vspace{30pt}
\appendix
\section{Sufficient conditions for the real ORF}
\label{AppA}
If the arm lengths of the two OLC detectors satisfy $L_1=L_2=L$, then $\beta_{12}=0$ and the ORF of the two OLC detectors can be written in the following form :
\begin{equation}
\Gamma_{12}(f)= {\displaystyle \sum_{A}} \int \frac{d^2 \hat{n}}{4 \pi} F_A(\hat{n},f)e^{i k \hat{n}\cdot\vec{m}_{12}},
\end{equation}
where $k=2\pi f/c$, and
\begin{equation}\label{eq3-58}
\begin{aligned}
F_A(\hat{n},f&)=(\hat{u}_1)^a (\hat{u}_1)^b {\bf{e}}^A_{ab}(\hat{n})(\hat{u}_2)^c (\hat{u}_2)^d {\bf{e}}^A_{cd}(\hat{n})\left(\frac{\pi f L}{c}\right)^2\\
&\times \mathrm{sinc} \Big[ \frac{\pi f L}{c} (1-\hat{u}_1\cdot\hat{n}) \Big] \mathrm{sinc} \Big[\frac{\pi f L}{c} (1-\hat{u}_2 \cdot \hat{n}) \Big],
\end{aligned}
\end{equation}
and
\begin{equation}
\vec{m}_{12}=\left(\vec{x}_2 -\frac{L}{2}\hat{u}_2\right)-
\left(\vec{x}_1-\frac{L}{2}\hat{u}_1\right).
\end{equation}

Furthermore, we define the vector:
\begin{equation}
\vec{s} = \hat{u}_1+\hat{u}_2.
\end{equation}
It can be shown that if $\vec{s}\cdot\vec{m}_{12}=0$, then the imaginary part of $\Gamma_{12}(f)$ is zero. In this case, the imaginary part of $\Gamma_{12}(f)$ can be written as:
\begin{equation}\label{eq3-61}
\operatorname{Im}\big[\Gamma_{12}(f)\big]={\displaystyle \sum_{A}} \int \frac{d^2 \hat{n}}{4 \pi} F_A(\hat{n},f) \sin( k\hat{n}\cdot\vec{m}_{12}).
\end{equation}
Consider a rotation $\mathcal{R}$ of $\pi$ around the axis $\vec{s}$. This is a measure-invariant orthogonal transformation satisfying:
\begin{equation}
d^2 (\mathcal{R}\hat{n})=d^2\hat{n}.
\end{equation}
Since the rotation axis is parallel to $\hat{u}_1+\hat{u}_2$, where $\hat{u}_1$ and $\hat{u}_2$ are unit vectors, it is straightforward to verify that
\begin{align}
\mathcal{R}\hat{u}_1=\hat{u}_2,\quad
\mathcal{R}\hat{u}_2=\hat{u}_1.\label{eq3-64}
\end{align}
Moreover, because $\vec{s}\cdot\vec{m}_{12}=0$, we have $\vec{m}_{12}\perp\vec{s}$. Rotating $\vec{m}_{12}$ by $\pi$ around the axis $\vec{s}$ reverses its direction:
\begin{equation}\label{eq3-65}
\mathcal{R}\vec{m}_{12}=-\vec{m}_{12}.
\end{equation}
From $\mathcal{R}=\mathcal{R}^{-1}$ and Eq.~\eqref{eq3-65}, we obtain:
\begin{equation}
\vec{m}_{12}\cdot(\mathcal{R}\hat{n})=(\mathcal{R}^{-1}\vec{m}_{12})\cdot\hat{n}=(\mathcal{R}\vec{m}_{12})\cdot\hat{n}=-\vec{m}_{12}\cdot\hat{n}.
\end{equation}
Hence,
\begin{equation}\label{eq3-67}
\sin(k\vec{m}_{12}\cdot\mathcal{R}\hat{n})=-\sin(k \vec{m}_{12}\cdot\hat{n}).
\end{equation}

Combining Eq.~\eqref{eq3-64} yields:
\begin{equation}
1-\hat{u}_1\cdot(\mathcal{R}\hat{n})=1-(\mathcal{R}^{-1}\hat{u}_1)\cdot\hat{n}=1-(\mathcal{R}\hat{u}_1)\cdot\hat{n}=1-\hat{u}_2\cdot\hat{n}.
\end{equation}
Similarly,
\begin{equation}
1-\hat{u}_2\cdot(\mathcal{R}\hat{n})=1-\hat{u}_1\cdot\hat{n}.
\end{equation}
Thus,
\begin{equation}\label{eq3-70}
\begin{aligned}
&\mathrm{sinc} \Big[ \frac{\pi f L}{c} (1-\hat{u}_1\cdot\mathcal{R}\hat{n}) \Big] \mathrm{sinc} \Big[\frac{\pi f L}{c} (1-\hat{u}_2 \cdot \mathcal{R}\hat{n}) \Big]\\
=\,&\mathrm{sinc} \Big[ \frac{\pi f L}{c} (1-\hat{u}_2\cdot\hat{n}) \Big] \mathrm{sinc} \Big[\frac{\pi f L}{c} (1-\hat{u}_1 \cdot \hat{n}) \Big].
\end{aligned}
\end{equation}

Denote
\begin{equation}
P_A(\hat{n})=(\hat{u}_1)^a (\hat{u}_1)^b {\bf{e}}^A_{ab}(\hat{n})(\hat{u}_2)^c (\hat{u}_2)^d {\bf{e}}^A_{cd}(\hat{n}).
\end{equation}
Since the rotation $\mathcal{R}$ is also an orthogonal transformation, we obtain:
\begin{equation}\label{eq3-72}
\begin{aligned}
P_A&(\mathcal{R}\hat{n})
=\,(\hat{u}_1)^a(\hat{u}_1)^b {\bf{e}}^A_{ab}(\mathcal{R}\hat{n})(\hat{u}_2)^c (\hat{u}_2)^d, {\bf{e}}^A_{cd}(\mathcal{R}\hat{n})\\
=\,&(\mathcal{R}^{-1}\hat{u}_1)^a (\mathcal{R}^{-1}\hat{u}_1)^b {\bf{e}}^A_{ab}(\hat{n})(\mathcal{R}^{-1}\hat{u}_2)^c(\mathcal{R}^{-1}\hat{u}_2)^d{\bf{e}}^A_{cd}(\hat{n})\\
=\,&(\mathcal{R}\hat{u}_1)^a (\mathcal{R}\hat{u}_1)^b {\bf{e}}^A_{ab}(\hat{n})(\mathcal{R}\hat{u}_2)^c(\mathcal{R}\hat{u}_2)^d{\bf{e}}^A_{cd}(\hat{n})\\
=\,&(\hat{u}_2)^a(\hat{u}_2)^b {\bf{e}}^A_{ab}(\hat{n})(\hat{u}_1)^c(\hat{u}_1)^d {\bf{e}}^A_{cd}(\hat{n})\\
=\,&P_A(\hat{n}).
\end{aligned}
\end{equation}

Combining Eqs.~\eqref{eq3-58}, \eqref{eq3-70} and \eqref{eq3-72}, we obtain:
\begin{equation}\label{eq3-73}
F_A(\mathcal{R}\hat{n},f)= F_A(\hat{n},f).
\end{equation}
Then, using Eqs.~\eqref{eq3-61}, \eqref{eq3-67} and \eqref{eq3-73}:
\begin{equation}
\begin{aligned}
{\displaystyle \sum_{A}} F_A(\mathcal{R}\hat{n},f)\sin(k \mathcal{R}\hat{n}\cdot\vec{m}_{12})\\
=-{\displaystyle \sum_{A}} F_A(\hat{n},f)\sin(k \hat{n}\cdot\vec{m}_{12}).
\end{aligned}
\end{equation}
Moreover, the transformation $\hat{n}\rightarrow\mathcal{R}\hat{n}$ preserves the spherical area measure:
\begin{equation}
\begin{aligned}
\operatorname{Im}\big[\Gamma_{12}(f)\big]=&{\displaystyle \sum_{A}} \int \frac{d^2 \hat{n}}{4 \pi} F_A(\hat{n},f) \sin( k \hat{n}\cdot\vec{m}{12})\\
=&{\displaystyle \sum_{A}} \int \frac{d^2 \hat{n}}{4 \pi} F_A(\mathcal{R}\hat{n},f) \sin( k \mathcal{R}\hat{n}\cdot\vec{m}_{12})\\
=&-\operatorname{Im}\left[\Gamma_{12}(f)\right].
\end{aligned}
\end{equation}
Therefore,
\begin{equation}
\operatorname{Im}\big[\Gamma_{12}(f)\big]=0.
\end{equation}

As for the real part of $\Gamma_{12}(f)$:
\begin{equation}
\cos(k\vec{m}_{12}\cdot\mathcal{R}\hat{n})=\cos(k \vec{m}_{12}\cdot\hat{n}),
\end{equation}
which is not necessarily zero. Hence, when $L_1=L_2$ and $\vec{s}\cdot\vec{m}_{12}=0$, the ORF $\Gamma_{12}(f)$ has only a real part, and its imaginary part must be zero.
 
\bibliography{ref2025.bib}

\begin{thebibliography}{45}%
\makeatletter
\providecommand \@ifxundefined [1]{%
 \@ifx{#1\undefined}
}%
\providecommand \@ifnum [1]{%
 \ifnum #1\expandafter \@firstoftwo
 \else \expandafter \@secondoftwo
 \fi
}%
\providecommand \@ifx [1]{%
 \ifx #1\expandafter \@firstoftwo
 \else \expandafter \@secondoftwo
 \fi
}%
\providecommand \natexlab [1]{#1}%
\providecommand \enquote  [1]{``#1''}%
\providecommand \bibnamefont  [1]{#1}%
\providecommand \bibfnamefont [1]{#1}%
\providecommand \citenamefont [1]{#1}%
\providecommand \href@noop [0]{\@secondoftwo}%
\providecommand \href [0]{\begingroup \@sanitize@url \@href}%
\providecommand \@href[1]{\@@startlink{#1}\@@href}%
\providecommand \@@href[1]{\endgroup#1\@@endlink}%
\providecommand \@sanitize@url [0]{\catcode `\\12\catcode `\$12\catcode
  `\&12\catcode `\#12\catcode `\^12\catcode `\_12\catcode `\%12\relax}%
\providecommand \@@startlink[1]{}%
\providecommand \@@endlink[0]{}%
\providecommand \url  [0]{\begingroup\@sanitize@url \@url }%
\providecommand \@url [1]{\endgroup\@href {#1}{\urlprefix }}%
\providecommand \urlprefix  [0]{URL }%
\providecommand \Eprint [0]{\href }%
\providecommand \doibase [0]{https://doi.org/}%
\providecommand \selectlanguage [0]{\@gobble}%
\providecommand \bibinfo  [0]{\@secondoftwo}%
\providecommand \bibfield  [0]{\@secondoftwo}%
\providecommand \translation [1]{[#1]}%
\providecommand \BibitemOpen [0]{}%
\providecommand \bibitemStop [0]{}%
\providecommand \bibitemNoStop [0]{.\EOS\space}%
\providecommand \EOS [0]{\spacefactor3000\relax}%
\providecommand \BibitemShut  [1]{\csname bibitem#1\endcsname}%
\let\auto@bib@innerbib\@empty
\bibitem [{\citenamefont {Abbott}\ \emph
  {et~al.}(2016{\natexlab{a}})\citenamefont {Abbott} \emph
  {et~al.}}]{LIGOScientific:2016aoc}%
  \BibitemOpen
  \bibfield  {author} {\bibinfo {author} {\bibfnamefont {B.~P.}\ \bibnamefont
  {Abbott}} \emph {et~al.} (\bibinfo {collaboration} {LIGO Scientific,
  Virgo}),\ }\bibfield  {title} {\bibinfo {title} {{Observation of
  Gravitational Waves from a Binary Black Hole Merger}},\ }\href
  {https://doi.org/10.1103/PhysRevLett.116.061102} {\bibfield  {journal}
  {\bibinfo  {journal} {Phys. Rev. Lett.}\ }\textbf {\bibinfo {volume} {116}},\
  \bibinfo {pages} {061102} (\bibinfo {year} {2016}{\natexlab{a}})},\ \Eprint
  {https://arxiv.org/abs/1602.03837} {arXiv:1602.03837 [gr-qc]} \BibitemShut
  {NoStop}%
\bibitem [{\citenamefont {Abbott}\ \emph {et~al.}(2017)\citenamefont {Abbott}
  \emph {et~al.}}]{LIGOScientific:2017vwq}%
  \BibitemOpen
  \bibfield  {author} {\bibinfo {author} {\bibfnamefont {B.~P.}\ \bibnamefont
  {Abbott}} \emph {et~al.} (\bibinfo {collaboration} {LIGO Scientific,
  Virgo}),\ }\bibfield  {title} {\bibinfo {title} {{GW170817: Observation of
  Gravitational Waves from a Binary Neutron Star Inspiral}},\ }\href
  {https://doi.org/10.1103/PhysRevLett.119.161101} {\bibfield  {journal}
  {\bibinfo  {journal} {Phys. Rev. Lett.}\ }\textbf {\bibinfo {volume} {119}},\
  \bibinfo {pages} {161101} (\bibinfo {year} {2017})},\ \Eprint
  {https://arxiv.org/abs/1710.05832} {arXiv:1710.05832 [gr-qc]} \BibitemShut
  {NoStop}%
\bibitem [{\citenamefont {Abbott}\ \emph
  {et~al.}(2016{\natexlab{b}})\citenamefont {Abbott} \emph
  {et~al.}}]{LIGOScientific:2016lio}%
  \BibitemOpen
  \bibfield  {author} {\bibinfo {author} {\bibfnamefont {B.~P.}\ \bibnamefont
  {Abbott}} \emph {et~al.} (\bibinfo {collaboration} {LIGO Scientific,
  Virgo}),\ }\bibfield  {title} {\bibinfo {title} {{Tests of general relativity
  with GW150914}},\ }\href {https://doi.org/10.1103/PhysRevLett.116.221101}
  {\bibfield  {journal} {\bibinfo  {journal} {Phys. Rev. Lett.}\ }\textbf
  {\bibinfo {volume} {116}},\ \bibinfo {pages} {221101} (\bibinfo {year}
  {2016}{\natexlab{b}})},\ \bibinfo {note} {[Erratum: Phys.Rev.Lett. 121,
  129902 (2018)]},\ \Eprint {https://arxiv.org/abs/1602.03841}
  {arXiv:1602.03841 [gr-qc]} \BibitemShut {NoStop}%
\bibitem [{\citenamefont {Abbott}\ \emph {et~al.}(2021)\citenamefont {Abbott}
  \emph {et~al.}}]{LIGOScientific:2021qlt}%
  \BibitemOpen
  \bibfield  {author} {\bibinfo {author} {\bibfnamefont {R.}~\bibnamefont
  {Abbott}} \emph {et~al.} (\bibinfo {collaboration} {LIGO Scientific, KAGRA,
  VIRGO}),\ }\bibfield  {title} {\bibinfo {title} {{Observation of
  Gravitational Waves from Two Neutron Star{\textendash}Black Hole
  Coalescences}},\ }\href {https://doi.org/10.3847/2041-8213/ac082e} {\bibfield
   {journal} {\bibinfo  {journal} {Astrophys. J. Lett.}\ }\textbf {\bibinfo
  {volume} {915}},\ \bibinfo {pages} {L5} (\bibinfo {year} {2021})},\ \Eprint
  {https://arxiv.org/abs/2106.15163} {arXiv:2106.15163 [astro-ph.HE]}
  \BibitemShut {NoStop}%
\bibitem [{\citenamefont {Aasi}\ \emph {et~al.}(2015)\citenamefont {Aasi} \emph
  {et~al.}}]{LIGOScientific:2014pky}%
  \BibitemOpen
  \bibfield  {author} {\bibinfo {author} {\bibfnamefont {J.}~\bibnamefont
  {Aasi}} \emph {et~al.} (\bibinfo {collaboration} {LIGO Scientific}),\
  }\bibfield  {title} {\bibinfo {title} {{Advanced LIGO}},\ }\href
  {https://doi.org/10.1088/0264-9381/32/7/074001} {\bibfield  {journal}
  {\bibinfo  {journal} {Class. Quant. Grav.}\ }\textbf {\bibinfo {volume}
  {32}},\ \bibinfo {pages} {074001} (\bibinfo {year} {2015})},\ \Eprint
  {https://arxiv.org/abs/1411.4547} {arXiv:1411.4547 [gr-qc]} \BibitemShut
  {NoStop}%
\bibitem [{\citenamefont {Acernese}\ \emph {et~al.}(2015)\citenamefont
  {Acernese} \emph {et~al.}}]{VIRGO:2014yos}%
  \BibitemOpen
  \bibfield  {author} {\bibinfo {author} {\bibfnamefont {F.}~\bibnamefont
  {Acernese}} \emph {et~al.} (\bibinfo {collaboration} {VIRGO}),\ }\bibfield
  {title} {\bibinfo {title} {{Advanced Virgo: a second-generation
  interferometric gravitational wave detector}},\ }\href
  {https://doi.org/10.1088/0264-9381/32/2/024001} {\bibfield  {journal}
  {\bibinfo  {journal} {Class. Quant. Grav.}\ }\textbf {\bibinfo {volume}
  {32}},\ \bibinfo {pages} {024001} (\bibinfo {year} {2015})},\ \Eprint
  {https://arxiv.org/abs/1408.3978} {arXiv:1408.3978 [gr-qc]} \BibitemShut
  {NoStop}%
\bibitem [{\citenamefont {Somiya}(2012)}]{Somiya:2011np}%
  \BibitemOpen
  \bibfield  {author} {\bibinfo {author} {\bibfnamefont {K.}~\bibnamefont
  {Somiya}} (\bibinfo {collaboration} {KAGRA}),\ }\bibfield  {title} {\bibinfo
  {title} {{Detector configuration of KAGRA: The Japanese cryogenic
  gravitational-wave detector}},\ }\href
  {https://doi.org/10.1088/0264-9381/29/12/124007} {\bibfield  {journal}
  {\bibinfo  {journal} {Class. Quant. Grav.}\ }\textbf {\bibinfo {volume}
  {29}},\ \bibinfo {pages} {124007} (\bibinfo {year} {2012})},\ \Eprint
  {https://arxiv.org/abs/1111.7185} {arXiv:1111.7185 [gr-qc]} \BibitemShut
  {NoStop}%
\bibitem [{\citenamefont {Punturo}\ \emph {et~al.}(2010)\citenamefont {Punturo}
  \emph {et~al.}}]{Punturo:2010zz}%
  \BibitemOpen
  \bibfield  {author} {\bibinfo {author} {\bibfnamefont {M.}~\bibnamefont
  {Punturo}} \emph {et~al.},\ }\bibfield  {title} {\bibinfo {title} {{The
  Einstein Telescope: A third-generation gravitational wave observatory}},\
  }\href {https://doi.org/10.1088/0264-9381/27/19/194002} {\bibfield  {journal}
  {\bibinfo  {journal} {Class. Quant. Grav.}\ }\textbf {\bibinfo {volume}
  {27}},\ \bibinfo {pages} {194002} (\bibinfo {year} {2010})}\BibitemShut
  {NoStop}%
\bibitem [{\citenamefont {Seoane}\ \emph {et~al.}(2013)\citenamefont {Seoane}
  \emph {et~al.}}]{eLISA:2013xep}%
  \BibitemOpen
  \bibfield  {author} {\bibinfo {author} {\bibfnamefont {P.~A.}\ \bibnamefont
  {Seoane}} \emph {et~al.} (\bibinfo {collaboration} {eLISA}),\ }\bibfield
  {title} {\bibinfo {title} {{The Gravitational Universe}},\ }\href@noop {}
  {\bibfield  {journal} {\bibinfo  {journal} {ArXiv}\ } (\bibinfo {year}
  {2013})},\ \Eprint {https://arxiv.org/abs/1305.5720} {arXiv:1305.5720
  [astro-ph.CO]} \BibitemShut {NoStop}%
\bibitem [{\citenamefont {Hu}\ and\ \citenamefont {Wu}(2017)}]{Hu:2017mde}%
  \BibitemOpen
  \bibfield  {author} {\bibinfo {author} {\bibfnamefont {W.-R.}\ \bibnamefont
  {Hu}}\ and\ \bibinfo {author} {\bibfnamefont {Y.-L.}\ \bibnamefont {Wu}},\
  }\bibfield  {title} {\bibinfo {title} {{The Taiji Program in Space for
  gravitational wave physics and the nature of gravity}},\ }\href
  {https://doi.org/10.1093/nsr/nwx116} {\bibfield  {journal} {\bibinfo
  {journal} {Natl. Sci. Rev.}\ }\textbf {\bibinfo {volume} {4}},\ \bibinfo
  {pages} {685} (\bibinfo {year} {2017})}\BibitemShut {NoStop}%
\bibitem [{\citenamefont {Luo}\ \emph {et~al.}(2016)\citenamefont {Luo} \emph
  {et~al.}}]{TianQin:2015yph}%
  \BibitemOpen
  \bibfield  {author} {\bibinfo {author} {\bibfnamefont {J.}~\bibnamefont
  {Luo}} \emph {et~al.} (\bibinfo {collaboration} {TianQin}),\ }\bibfield
  {title} {\bibinfo {title} {{TianQin: a space-borne gravitational wave
  detector}},\ }\href {https://doi.org/10.1088/0264-9381/33/3/035010}
  {\bibfield  {journal} {\bibinfo  {journal} {Class. Quant. Grav.}\ }\textbf
  {\bibinfo {volume} {33}},\ \bibinfo {pages} {035010} (\bibinfo {year}
  {2016})},\ \Eprint {https://arxiv.org/abs/1512.02076} {arXiv:1512.02076
  [astro-ph.IM]} \BibitemShut {NoStop}%
\bibitem [{\citenamefont {Detweiler}(1979)}]{Detweiler:1979wn}%
  \BibitemOpen
  \bibfield  {author} {\bibinfo {author} {\bibfnamefont {S.~L.}\ \bibnamefont
  {Detweiler}},\ }\bibfield  {title} {\bibinfo {title} {{Pulsar timing
  measurements and the search for gravitational waves}},\ }\href
  {https://doi.org/10.1086/157593} {\bibfield  {journal} {\bibinfo  {journal}
  {Astrophys. J.}\ }\textbf {\bibinfo {volume} {234}},\ \bibinfo {pages} {1100}
  (\bibinfo {year} {1979})}\BibitemShut {NoStop}%
\bibitem [{\citenamefont {Burke-Spolaor}\ \emph {et~al.}(2019)\citenamefont
  {Burke-Spolaor} \emph {et~al.}}]{Burke-Spolaor:2018bvk}%
  \BibitemOpen
  \bibfield  {author} {\bibinfo {author} {\bibfnamefont {S.}~\bibnamefont
  {Burke-Spolaor}} \emph {et~al.},\ }\bibfield  {title} {\bibinfo {title} {{The
  Astrophysics of Nanohertz Gravitational Waves}},\ }\href
  {https://doi.org/10.1007/s00159-019-0115-7} {\bibfield  {journal} {\bibinfo
  {journal} {Astron. Astrophys. Rev.}\ }\textbf {\bibinfo {volume} {27}},\
  \bibinfo {pages} {5} (\bibinfo {year} {2019})},\ \Eprint
  {https://arxiv.org/abs/1811.08826} {arXiv:1811.08826 [astro-ph.HE]}
  \BibitemShut {NoStop}%
\bibitem [{\citenamefont {Arzoumanian}\ \emph {et~al.}(2020)\citenamefont
  {Arzoumanian} \emph {et~al.}}]{NANOGrav:2020bcs}%
  \BibitemOpen
  \bibfield  {author} {\bibinfo {author} {\bibfnamefont {Z.}~\bibnamefont
  {Arzoumanian}} \emph {et~al.} (\bibinfo {collaboration} {NANOGrav}),\
  }\bibfield  {title} {\bibinfo {title} {{The NANOGrav 12.5 yr Data Set: Search
  for an Isotropic Stochastic Gravitational-wave Background}},\ }\href
  {https://doi.org/10.3847/2041-8213/abd401} {\bibfield  {journal} {\bibinfo
  {journal} {Astrophys. J. Lett.}\ }\textbf {\bibinfo {volume} {905}},\
  \bibinfo {pages} {L34} (\bibinfo {year} {2020})},\ \Eprint
  {https://arxiv.org/abs/2009.04496} {arXiv:2009.04496 [astro-ph.HE]}
  \BibitemShut {NoStop}%
\bibitem [{\citenamefont {Goncharov}\ \emph {et~al.}(2021)\citenamefont
  {Goncharov} \emph {et~al.}}]{Goncharov:2021oub}%
  \BibitemOpen
  \bibfield  {author} {\bibinfo {author} {\bibfnamefont {B.}~\bibnamefont
  {Goncharov}} \emph {et~al.},\ }\bibfield  {title} {\bibinfo {title} {{On the
  Evidence for a Common-spectrum Process in the Search for the Nanohertz
  Gravitational-wave Background with the Parkes Pulsar Timing Array}},\ }\href
  {https://doi.org/10.3847/2041-8213/ac17f4} {\bibfield  {journal} {\bibinfo
  {journal} {Astrophys. J. Lett.}\ }\textbf {\bibinfo {volume} {917}},\
  \bibinfo {pages} {L19} (\bibinfo {year} {2021})},\ \Eprint
  {https://arxiv.org/abs/2107.12112} {arXiv:2107.12112 [astro-ph.HE]}
  \BibitemShut {NoStop}%
\bibitem [{\citenamefont {Chen}\ \emph {et~al.}(2021)\citenamefont {Chen} \emph
  {et~al.}}]{EPTA:2021crs}%
  \BibitemOpen
  \bibfield  {author} {\bibinfo {author} {\bibfnamefont {S.}~\bibnamefont
  {Chen}} \emph {et~al.} (\bibinfo {collaboration} {EPTA}),\ }\bibfield
  {title} {\bibinfo {title} {{Common-red-signal analysis with 24-yr
  high-precision timing of the European Pulsar Timing Array: inferences in the
  stochastic gravitational-wave background search}},\ }\href
  {https://doi.org/10.1093/mnras/stab2833} {\bibfield  {journal} {\bibinfo
  {journal} {Mon. Not. Roy. Astron. Soc.}\ }\textbf {\bibinfo {volume} {508}},\
  \bibinfo {pages} {4970} (\bibinfo {year} {2021})},\ \Eprint
  {https://arxiv.org/abs/2110.13184} {arXiv:2110.13184 [astro-ph.HE]}
  \BibitemShut {NoStop}%
\bibitem [{\citenamefont {Antoniadis}\ \emph {et~al.}(2022)\citenamefont
  {Antoniadis} \emph {et~al.}}]{Antoniadis:2022pcn}%
  \BibitemOpen
  \bibfield  {author} {\bibinfo {author} {\bibfnamefont {J.}~\bibnamefont
  {Antoniadis}} \emph {et~al.},\ }\bibfield  {title} {\bibinfo {title} {{The
  International Pulsar Timing Array second data release: Search for an
  isotropic gravitational wave background}},\ }\href
  {https://doi.org/10.1093/mnras/stab3418} {\bibfield  {journal} {\bibinfo
  {journal} {Mon. Not. Roy. Astron. Soc.}\ }\textbf {\bibinfo {volume} {510}},\
  \bibinfo {pages} {4873} (\bibinfo {year} {2022})},\ \Eprint
  {https://arxiv.org/abs/2201.03980} {arXiv:2201.03980 [astro-ph.HE]}
  \BibitemShut {NoStop}%
\bibitem [{\citenamefont {He}\ and\ \citenamefont {Zhang}(2020)}]{He:2020elt}%
  \BibitemOpen
  \bibfield  {author} {\bibinfo {author} {\bibfnamefont {F.}~\bibnamefont
  {He}}\ and\ \bibinfo {author} {\bibfnamefont {B.}~\bibnamefont {Zhang}},\
  }\bibfield  {title} {\bibinfo {title} {{A protocol of potential advantage in
  the low frequency range to gravitational wave detection with space based
  optical atomic clocks}},\ }\href
  {https://doi.org/10.1140/epjd/e2020-100611-y} {\bibfield  {journal} {\bibinfo
   {journal} {Eur. Phys. J. D}\ }\textbf {\bibinfo {volume} {74}},\ \bibinfo
  {pages} {94} (\bibinfo {year} {2020})},\ \Eprint
  {https://arxiv.org/abs/2005.06817} {arXiv:2005.06817 [gr-qc]} \BibitemShut
  {NoStop}%
\bibitem [{\citenamefont {Ebisuzaki}\ \emph {et~al.}(2019)\citenamefont
  {Ebisuzaki}, \citenamefont {Katori}, \citenamefont {Makino}, \citenamefont
  {Noda}, \citenamefont {Shinkai},\ and\ \citenamefont
  {Tamagawa}}]{Ebisuzaki:2018ujm}%
  \BibitemOpen
  \bibfield  {author} {\bibinfo {author} {\bibfnamefont {T.}~\bibnamefont
  {Ebisuzaki}}, \bibinfo {author} {\bibfnamefont {H.}~\bibnamefont {Katori}},
  \bibinfo {author} {\bibfnamefont {J.}~\bibnamefont {Makino}}, \bibinfo
  {author} {\bibfnamefont {A.}~\bibnamefont {Noda}}, \bibinfo {author}
  {\bibfnamefont {H.}~\bibnamefont {Shinkai}},\ and\ \bibinfo {author}
  {\bibfnamefont {T.}~\bibnamefont {Tamagawa}},\ }\bibfield  {title} {\bibinfo
  {title} {{INO: Interplanetary Network of Optical Lattice Clocks}},\ }\href
  {https://doi.org/10.1142/S0218271819400029} {\bibfield  {journal} {\bibinfo
  {journal} {Int. J. Mod. Phys. D}\ }\textbf {\bibinfo {volume} {29}},\
  \bibinfo {pages} {1940002} (\bibinfo {year} {2019})},\ \Eprint
  {https://arxiv.org/abs/1809.10317} {arXiv:1809.10317 [astro-ph.IM]}
  \BibitemShut {NoStop}%
\bibitem [{\citenamefont {Hinkley}\ \emph {et~al.}(2013)\citenamefont
  {Hinkley}, \citenamefont {Sherman}, \citenamefont {Phillips}, \citenamefont
  {Schioppo}, \citenamefont {Lemke}, \citenamefont {Beloy}, \citenamefont
  {Pizzocaro}, \citenamefont {Oates},\ and\ \citenamefont
  {Ludlow}}]{Hinkley:2013oos}%
  \BibitemOpen
  \bibfield  {author} {\bibinfo {author} {\bibfnamefont {N.}~\bibnamefont
  {Hinkley}}, \bibinfo {author} {\bibfnamefont {J.~A.}\ \bibnamefont
  {Sherman}}, \bibinfo {author} {\bibfnamefont {N.~B.}\ \bibnamefont
  {Phillips}}, \bibinfo {author} {\bibfnamefont {M.}~\bibnamefont {Schioppo}},
  \bibinfo {author} {\bibfnamefont {N.~D.}\ \bibnamefont {Lemke}}, \bibinfo
  {author} {\bibfnamefont {K.}~\bibnamefont {Beloy}}, \bibinfo {author}
  {\bibfnamefont {M.}~\bibnamefont {Pizzocaro}}, \bibinfo {author}
  {\bibfnamefont {C.~W.}\ \bibnamefont {Oates}},\ and\ \bibinfo {author}
  {\bibfnamefont {A.~D.}\ \bibnamefont {Ludlow}},\ }\bibfield  {title}
  {\bibinfo {title} {{An Atomic Clock with 10{\textendash}18 Instability}},\
  }\href {https://doi.org/10.1126/science.1240420} {\bibfield  {journal}
  {\bibinfo  {journal} {Science}\ }\textbf {\bibinfo {volume} {341}},\ \bibinfo
  {pages} {1215} (\bibinfo {year} {2013})},\ \Eprint
  {https://arxiv.org/abs/1305.5869} {arXiv:1305.5869 [physics.atom-ph]}
  \BibitemShut {NoStop}%
\bibitem [{\citenamefont {Bloom}\ \emph {et~al.}(2014)\citenamefont {Bloom},
  \citenamefont {Nicholson}, \citenamefont {Williams}, \citenamefont
  {Campbell}, \citenamefont {Bishof}, \citenamefont {Zhang}, \citenamefont
  {Zhang}, \citenamefont {Bromley},\ and\ \citenamefont {Ye}}]{Bloom:2013uoa}%
  \BibitemOpen
  \bibfield  {author} {\bibinfo {author} {\bibfnamefont {B.~J.}\ \bibnamefont
  {Bloom}}, \bibinfo {author} {\bibfnamefont {T.~L.}\ \bibnamefont
  {Nicholson}}, \bibinfo {author} {\bibfnamefont {J.~R.}\ \bibnamefont
  {Williams}}, \bibinfo {author} {\bibfnamefont {S.~L.}\ \bibnamefont
  {Campbell}}, \bibinfo {author} {\bibfnamefont {M.}~\bibnamefont {Bishof}},
  \bibinfo {author} {\bibfnamefont {X.}~\bibnamefont {Zhang}}, \bibinfo
  {author} {\bibfnamefont {W.}~\bibnamefont {Zhang}}, \bibinfo {author}
  {\bibfnamefont {S.~L.}\ \bibnamefont {Bromley}},\ and\ \bibinfo {author}
  {\bibfnamefont {J.}~\bibnamefont {Ye}},\ }\bibfield  {title} {\bibinfo
  {title} {{An Optical Lattice Clock with Accuracy and Stability at the
  $10^{-18}$ Level}},\ }\href {https://doi.org/10.1038/nature12941} {\bibfield
  {journal} {\bibinfo  {journal} {Nature}\ }\textbf {\bibinfo {volume} {506}},\
  \bibinfo {pages} {71} (\bibinfo {year} {2014})},\ \Eprint
  {https://arxiv.org/abs/1309.1137} {arXiv:1309.1137 [physics.atom-ph]}
  \BibitemShut {NoStop}%
\bibitem [{\citenamefont {Kolkowitz}\ \emph {et~al.}(2016)\citenamefont
  {Kolkowitz}, \citenamefont {Pikovski}, \citenamefont {Langellier},
  \citenamefont {Lukin}, \citenamefont {Walsworth},\ and\ \citenamefont
  {Ye}}]{Kolkowitz:2016wyg}%
  \BibitemOpen
  \bibfield  {author} {\bibinfo {author} {\bibfnamefont {S.}~\bibnamefont
  {Kolkowitz}}, \bibinfo {author} {\bibfnamefont {I.}~\bibnamefont {Pikovski}},
  \bibinfo {author} {\bibfnamefont {N.}~\bibnamefont {Langellier}}, \bibinfo
  {author} {\bibfnamefont {M.~D.}\ \bibnamefont {Lukin}}, \bibinfo {author}
  {\bibfnamefont {R.~L.}\ \bibnamefont {Walsworth}},\ and\ \bibinfo {author}
  {\bibfnamefont {J.}~\bibnamefont {Ye}},\ }\bibfield  {title} {\bibinfo
  {title} {{Gravitational wave detection with optical lattice atomic clocks}},\
  }\href {https://doi.org/10.1103/PhysRevD.94.124043} {\bibfield  {journal}
  {\bibinfo  {journal} {Phys. Rev. D}\ }\textbf {\bibinfo {volume} {94}},\
  \bibinfo {pages} {124043} (\bibinfo {year} {2016})},\ \Eprint
  {https://arxiv.org/abs/1606.01859} {arXiv:1606.01859 [physics.atom-ph]}
  \BibitemShut {NoStop}%
\bibitem [{\citenamefont {Christensen}(2019)}]{Christensen:2018iqi}%
  \BibitemOpen
  \bibfield  {author} {\bibinfo {author} {\bibfnamefont {N.}~\bibnamefont
  {Christensen}},\ }\bibfield  {title} {\bibinfo {title} {{Stochastic
  Gravitational Wave Backgrounds}},\ }\href
  {https://doi.org/10.1088/1361-6633/aae6b5} {\bibfield  {journal} {\bibinfo
  {journal} {Rept. Prog. Phys.}\ }\textbf {\bibinfo {volume} {82}},\ \bibinfo
  {pages} {016903} (\bibinfo {year} {2019})},\ \Eprint
  {https://arxiv.org/abs/1811.08797} {arXiv:1811.08797 [gr-qc]} \BibitemShut
  {NoStop}%
\bibitem [{\citenamefont {Maggiore}(2000)}]{Maggiore:1999vm}%
  \BibitemOpen
  \bibfield  {author} {\bibinfo {author} {\bibfnamefont {M.}~\bibnamefont
  {Maggiore}},\ }\bibfield  {title} {\bibinfo {title} {{Gravitational wave
  experiments and early universe cosmology}},\ }\href
  {https://doi.org/10.1016/S0370-1573(99)00102-7} {\bibfield  {journal}
  {\bibinfo  {journal} {Phys. Rept.}\ }\textbf {\bibinfo {volume} {331}},\
  \bibinfo {pages} {283} (\bibinfo {year} {2000})},\ \Eprint
  {https://arxiv.org/abs/gr-qc/9909001} {arXiv:gr-qc/9909001} \BibitemShut
  {NoStop}%
\bibitem [{\citenamefont {Caprini}\ and\ \citenamefont
  {Figueroa}(2018)}]{Caprini:2018mtu}%
  \BibitemOpen
  \bibfield  {author} {\bibinfo {author} {\bibfnamefont {C.}~\bibnamefont
  {Caprini}}\ and\ \bibinfo {author} {\bibfnamefont {D.~G.}\ \bibnamefont
  {Figueroa}},\ }\bibfield  {title} {\bibinfo {title} {{Cosmological
  Backgrounds of Gravitational Waves}},\ }\href
  {https://doi.org/10.1088/1361-6382/aac608} {\bibfield  {journal} {\bibinfo
  {journal} {Class. Quant. Grav.}\ }\textbf {\bibinfo {volume} {35}},\ \bibinfo
  {pages} {163001} (\bibinfo {year} {2018})},\ \Eprint
  {https://arxiv.org/abs/1801.04268} {arXiv:1801.04268 [astro-ph.CO]}
  \BibitemShut {NoStop}%
\bibitem [{\citenamefont {Damour}\ and\ \citenamefont
  {Vilenkin}(2001)}]{Damour:2001bk}%
  \BibitemOpen
  \bibfield  {author} {\bibinfo {author} {\bibfnamefont {T.}~\bibnamefont
  {Damour}}\ and\ \bibinfo {author} {\bibfnamefont {A.}~\bibnamefont
  {Vilenkin}},\ }\bibfield  {title} {\bibinfo {title} {{Gravitational wave
  bursts from cusps and kinks on cosmic strings}},\ }\href
  {https://doi.org/10.1103/PhysRevD.64.064008} {\bibfield  {journal} {\bibinfo
  {journal} {Phys. Rev. D}\ }\textbf {\bibinfo {volume} {64}},\ \bibinfo
  {pages} {064008} (\bibinfo {year} {2001})},\ \Eprint
  {https://arxiv.org/abs/gr-qc/0104026} {arXiv:gr-qc/0104026} \BibitemShut
  {NoStop}%
\bibitem [{\citenamefont {Siemens}\ \emph {et~al.}(2007)\citenamefont
  {Siemens}, \citenamefont {Mandic},\ and\ \citenamefont
  {Creighton}}]{Siemens:2006yp}%
  \BibitemOpen
  \bibfield  {author} {\bibinfo {author} {\bibfnamefont {X.}~\bibnamefont
  {Siemens}}, \bibinfo {author} {\bibfnamefont {V.}~\bibnamefont {Mandic}},\
  and\ \bibinfo {author} {\bibfnamefont {J.}~\bibnamefont {Creighton}},\
  }\bibfield  {title} {\bibinfo {title} {{Gravitational wave stochastic
  background from cosmic (super)strings}},\ }\href
  {https://doi.org/10.1103/PhysRevLett.98.111101} {\bibfield  {journal}
  {\bibinfo  {journal} {Phys. Rev. Lett.}\ }\textbf {\bibinfo {volume} {98}},\
  \bibinfo {pages} {111101} (\bibinfo {year} {2007})},\ \Eprint
  {https://arxiv.org/abs/astro-ph/0610920} {arXiv:astro-ph/0610920}
  \BibitemShut {NoStop}%
\bibitem [{\citenamefont {Witten}(1984)}]{Witten:1984rs}%
  \BibitemOpen
  \bibfield  {author} {\bibinfo {author} {\bibfnamefont {E.}~\bibnamefont
  {Witten}},\ }\bibfield  {title} {\bibinfo {title} {{Cosmic Separation of
  Phases}},\ }\href {https://doi.org/10.1103/PhysRevD.30.272} {\bibfield
  {journal} {\bibinfo  {journal} {Phys. Rev. D}\ }\textbf {\bibinfo {volume}
  {30}},\ \bibinfo {pages} {272} (\bibinfo {year} {1984})}\BibitemShut
  {NoStop}%
\bibitem [{\citenamefont {Kamionkowski}\ \emph {et~al.}(1994)\citenamefont
  {Kamionkowski}, \citenamefont {Kosowsky},\ and\ \citenamefont
  {Turner}}]{Kamionkowski:1993fg}%
  \BibitemOpen
  \bibfield  {author} {\bibinfo {author} {\bibfnamefont {M.}~\bibnamefont
  {Kamionkowski}}, \bibinfo {author} {\bibfnamefont {A.}~\bibnamefont
  {Kosowsky}},\ and\ \bibinfo {author} {\bibfnamefont {M.~S.}\ \bibnamefont
  {Turner}},\ }\bibfield  {title} {\bibinfo {title} {{Gravitational radiation
  from first order phase transitions}},\ }\href
  {https://doi.org/10.1103/PhysRevD.49.2837} {\bibfield  {journal} {\bibinfo
  {journal} {Phys. Rev. D}\ }\textbf {\bibinfo {volume} {49}},\ \bibinfo
  {pages} {2837} (\bibinfo {year} {1994})},\ \Eprint
  {https://arxiv.org/abs/astro-ph/9310044} {arXiv:astro-ph/9310044}
  \BibitemShut {NoStop}%
\bibitem [{\citenamefont {Mukhanov}\ \emph {et~al.}(1992)\citenamefont
  {Mukhanov}, \citenamefont {Feldman},\ and\ \citenamefont
  {Brandenberger}}]{Mukhanov:1990me}%
  \BibitemOpen
  \bibfield  {author} {\bibinfo {author} {\bibfnamefont {V.~F.}\ \bibnamefont
  {Mukhanov}}, \bibinfo {author} {\bibfnamefont {H.~A.}\ \bibnamefont
  {Feldman}},\ and\ \bibinfo {author} {\bibfnamefont {R.~H.}\ \bibnamefont
  {Brandenberger}},\ }\bibfield  {title} {\bibinfo {title} {{Theory of
  cosmological perturbations. Part 1. Classical perturbations. Part 2. Quantum
  theory of perturbations. Part 3. Extensions}},\ }\href
  {https://doi.org/10.1016/0370-1573(92)90044-Z} {\bibfield  {journal}
  {\bibinfo  {journal} {Phys. Rept.}\ }\textbf {\bibinfo {volume} {215}},\
  \bibinfo {pages} {203} (\bibinfo {year} {1992})}\BibitemShut {NoStop}%
\bibitem [{\citenamefont {Turner}(1997)}]{Turner:1996ck}%
  \BibitemOpen
  \bibfield  {author} {\bibinfo {author} {\bibfnamefont {M.~S.}\ \bibnamefont
  {Turner}},\ }\bibfield  {title} {\bibinfo {title} {{Detectability of
  inflation produced gravitational waves}},\ }\href
  {https://doi.org/10.1103/PhysRevD.55.R435} {\bibfield  {journal} {\bibinfo
  {journal} {Phys. Rev. D}\ }\textbf {\bibinfo {volume} {55}},\ \bibinfo
  {pages} {R435} (\bibinfo {year} {1997})},\ \Eprint
  {https://arxiv.org/abs/astro-ph/9607066} {arXiv:astro-ph/9607066}
  \BibitemShut {NoStop}%
\bibitem [{\citenamefont {Farmer}\ and\ \citenamefont
  {Phinney}(2003)}]{Farmer:2003pa}%
  \BibitemOpen
  \bibfield  {author} {\bibinfo {author} {\bibfnamefont {A.~J.}\ \bibnamefont
  {Farmer}}\ and\ \bibinfo {author} {\bibfnamefont {E.~S.}\ \bibnamefont
  {Phinney}},\ }\bibfield  {title} {\bibinfo {title} {{The gravitational wave
  background from cosmological compact binaries}},\ }\href
  {https://doi.org/10.1111/j.1365-2966.2003.07176.x} {\bibfield  {journal}
  {\bibinfo  {journal} {Mon. Not. Roy. Astron. Soc.}\ }\textbf {\bibinfo
  {volume} {346}},\ \bibinfo {pages} {1197} (\bibinfo {year} {2003})},\ \Eprint
  {https://arxiv.org/abs/astro-ph/0304393} {arXiv:astro-ph/0304393}
  \BibitemShut {NoStop}%
\bibitem [{\citenamefont {Taylor}\ and\ \citenamefont
  {Gair}(2012)}]{Taylor:2012db}%
  \BibitemOpen
  \bibfield  {author} {\bibinfo {author} {\bibfnamefont {S.~R.}\ \bibnamefont
  {Taylor}}\ and\ \bibinfo {author} {\bibfnamefont {J.~R.}\ \bibnamefont
  {Gair}},\ }\bibfield  {title} {\bibinfo {title} {{Cosmology with the lights
  off: standard sirens in the Einstein Telescope era}},\ }\href
  {https://doi.org/10.1103/PhysRevD.86.023502} {\bibfield  {journal} {\bibinfo
  {journal} {Phys. Rev. D}\ }\textbf {\bibinfo {volume} {86}},\ \bibinfo
  {pages} {023502} (\bibinfo {year} {2012})},\ \Eprint
  {https://arxiv.org/abs/1204.6739} {arXiv:1204.6739 [astro-ph.CO]}
  \BibitemShut {NoStop}%
\bibitem [{\citenamefont {Allen}\ and\ \citenamefont
  {Romano}(1999)}]{Allen:1997ad}%
  \BibitemOpen
  \bibfield  {author} {\bibinfo {author} {\bibfnamefont {B.}~\bibnamefont
  {Allen}}\ and\ \bibinfo {author} {\bibfnamefont {J.~D.}\ \bibnamefont
  {Romano}},\ }\bibfield  {title} {\bibinfo {title} {{Detecting a stochastic
  background of gravitational radiation: Signal processing strategies and
  sensitivities}},\ }\href {https://doi.org/10.1103/PhysRevD.59.102001}
  {\bibfield  {journal} {\bibinfo  {journal} {Phys. Rev. D}\ }\textbf {\bibinfo
  {volume} {59}},\ \bibinfo {pages} {102001} (\bibinfo {year} {1999})},\
  \Eprint {https://arxiv.org/abs/gr-qc/9710117} {arXiv:gr-qc/9710117}
  \BibitemShut {NoStop}%
\bibitem [{\citenamefont {Thrane}\ and\ \citenamefont
  {Romano}(2013)}]{Thrane:2013oya}%
  \BibitemOpen
  \bibfield  {author} {\bibinfo {author} {\bibfnamefont {E.}~\bibnamefont
  {Thrane}}\ and\ \bibinfo {author} {\bibfnamefont {J.~D.}\ \bibnamefont
  {Romano}},\ }\bibfield  {title} {\bibinfo {title} {{Sensitivity curves for
  searches for gravitational-wave backgrounds}},\ }\href
  {https://doi.org/10.1103/PhysRevD.88.124032} {\bibfield  {journal} {\bibinfo
  {journal} {Phys. Rev. D}\ }\textbf {\bibinfo {volume} {88}},\ \bibinfo
  {pages} {124032} (\bibinfo {year} {2013})},\ \Eprint
  {https://arxiv.org/abs/1310.5300} {arXiv:1310.5300 [astro-ph.IM]}
  \BibitemShut {NoStop}%
\bibitem [{\citenamefont {Wang}\ \emph {et~al.}(2025)\citenamefont {Wang},
  \citenamefont {Li}, \citenamefont {Xiao}, \citenamefont {Mo},\ and\
  \citenamefont {Cai}}]{Wang:2024tnk}%
  \BibitemOpen
  \bibfield  {author} {\bibinfo {author} {\bibfnamefont {B.}~\bibnamefont
  {Wang}}, \bibinfo {author} {\bibfnamefont {B.}~\bibnamefont {Li}}, \bibinfo
  {author} {\bibfnamefont {Q.}~\bibnamefont {Xiao}}, \bibinfo {author}
  {\bibfnamefont {G.}~\bibnamefont {Mo}},\ and\ \bibinfo {author}
  {\bibfnamefont {Y.-F.}\ \bibnamefont {Cai}},\ }\bibfield  {title} {\bibinfo
  {title} {{Space-based optical lattice clocks as gravitational wave detectors
  in search for new physics}},\ }\href
  {https://doi.org/10.1007/s11433-024-2573-3} {\bibfield  {journal} {\bibinfo
  {journal} {Sci. China Phys. Mech. Astron.}\ }\textbf {\bibinfo {volume}
  {68}},\ \bibinfo {pages} {249512} (\bibinfo {year} {2025})},\ \Eprint
  {https://arxiv.org/abs/2410.04340} {arXiv:2410.04340 [gr-qc]} \BibitemShut
  {NoStop}%
\bibitem [{\citenamefont {Hu}\ \emph {et~al.}(2025)\citenamefont {Hu},
  \citenamefont {Wang}, \citenamefont {Tan},\ and\ \citenamefont
  {Shao}}]{Hu:2025fev}%
  \BibitemOpen
  \bibfield  {author} {\bibinfo {author} {\bibfnamefont {Y.}~\bibnamefont
  {Hu}}, \bibinfo {author} {\bibfnamefont {P.-P.}\ \bibnamefont {Wang}},
  \bibinfo {author} {\bibfnamefont {Y.-J.}\ \bibnamefont {Tan}},\ and\ \bibinfo
  {author} {\bibfnamefont {C.-G.}\ \bibnamefont {Shao}},\ }\bibfield  {title}
  {\bibinfo {title} {{Universal calculation approach of overlap reduction
  function for pulsar timing array and laser interferometer detector}},\ }\href
  {https://doi.org/10.1103/PhysRevD.111.084065} {\bibfield  {journal} {\bibinfo
   {journal} {Phys. Rev. D}\ }\textbf {\bibinfo {volume} {111}},\ \bibinfo
  {pages} {084065} (\bibinfo {year} {2025})}\BibitemShut {NoStop}%
\bibitem [{\citenamefont {Hu}\ \emph {et~al.}(2022)\citenamefont {Hu},
  \citenamefont {Wang}, \citenamefont {Tan},\ and\ \citenamefont
  {Shao}}]{Hu:2022ujx}%
  \BibitemOpen
  \bibfield  {author} {\bibinfo {author} {\bibfnamefont {Y.}~\bibnamefont
  {Hu}}, \bibinfo {author} {\bibfnamefont {P.-P.}\ \bibnamefont {Wang}},
  \bibinfo {author} {\bibfnamefont {Y.-J.}\ \bibnamefont {Tan}},\ and\ \bibinfo
  {author} {\bibfnamefont {C.-G.}\ \bibnamefont {Shao}},\ }\bibfield  {title}
  {\bibinfo {title} {{Full analytic expression of overlap reduction function
  for gravitational wave background with pulsar timing arrays}},\ }\href
  {https://doi.org/10.1103/PhysRevD.106.024005} {\bibfield  {journal} {\bibinfo
   {journal} {Phys. Rev. D}\ }\textbf {\bibinfo {volume} {106}},\ \bibinfo
  {pages} {024005} (\bibinfo {year} {2022})},\ \Eprint
  {https://arxiv.org/abs/2205.09272} {arXiv:2205.09272 [gr-qc]} \BibitemShut
  {NoStop}%
\bibitem [{\citenamefont {Hellings}\ and\ \citenamefont
  {Downs}(1983)}]{Hellings:1983fr}%
  \BibitemOpen
  \bibfield  {author} {\bibinfo {author} {\bibfnamefont {R.~w.}\ \bibnamefont
  {Hellings}}\ and\ \bibinfo {author} {\bibfnamefont {G.~s.}\ \bibnamefont
  {Downs}},\ }\bibfield  {title} {\bibinfo {title} {{UPPER LIMITS ON THE
  ISOTROPIC GRAVITATIONAL RADIATION BACKGROUND FROM PULSAR TIMING ANALYSIS}},\
  }\href {https://doi.org/10.1086/183954} {\bibfield  {journal} {\bibinfo
  {journal} {Astrophys. J. Lett.}\ }\textbf {\bibinfo {volume} {265}},\
  \bibinfo {pages} {L39} (\bibinfo {year} {1983})}\BibitemShut {NoStop}%
\bibitem [{\citenamefont {Cornish}\ and\ \citenamefont
  {Larson}(2001)}]{Cornish:2001qi}%
  \BibitemOpen
  \bibfield  {author} {\bibinfo {author} {\bibfnamefont {N.~J.}\ \bibnamefont
  {Cornish}}\ and\ \bibinfo {author} {\bibfnamefont {S.~L.}\ \bibnamefont
  {Larson}},\ }\bibfield  {title} {\bibinfo {title} {{Space missions to detect
  the cosmic gravitational wave background}},\ }\href
  {https://doi.org/10.1088/0264-9381/18/17/308} {\bibfield  {journal} {\bibinfo
   {journal} {Class. Quant. Grav.}\ }\textbf {\bibinfo {volume} {18}},\
  \bibinfo {pages} {3473} (\bibinfo {year} {2001})},\ \Eprint
  {https://arxiv.org/abs/gr-qc/0103075} {arXiv:gr-qc/0103075} \BibitemShut
  {NoStop}%
\bibitem [{\citenamefont {Robson}\ \emph {et~al.}(2019)\citenamefont {Robson},
  \citenamefont {Cornish},\ and\ \citenamefont {Liu}}]{Robson:2018ifk}%
  \BibitemOpen
  \bibfield  {author} {\bibinfo {author} {\bibfnamefont {T.}~\bibnamefont
  {Robson}}, \bibinfo {author} {\bibfnamefont {N.~J.}\ \bibnamefont
  {Cornish}},\ and\ \bibinfo {author} {\bibfnamefont {C.}~\bibnamefont {Liu}},\
  }\bibfield  {title} {\bibinfo {title} {{The construction and use of LISA
  sensitivity curves}},\ }\href {https://doi.org/10.1088/1361-6382/ab1101}
  {\bibfield  {journal} {\bibinfo  {journal} {Class. Quant. Grav.}\ }\textbf
  {\bibinfo {volume} {36}},\ \bibinfo {pages} {105011} (\bibinfo {year}
  {2019})},\ \Eprint {https://arxiv.org/abs/1803.01944} {arXiv:1803.01944
  [astro-ph.HE]} \BibitemShut {NoStop}%
\bibitem [{\citenamefont {Luo}\ \emph {et~al.}(2021)\citenamefont {Luo},
  \citenamefont {Wang}, \citenamefont {Wu}, \citenamefont {Hu},\ and\
  \citenamefont {Jin}}]{Luo:2021qji}%
  \BibitemOpen
  \bibfield  {author} {\bibinfo {author} {\bibfnamefont {Z.}~\bibnamefont
  {Luo}}, \bibinfo {author} {\bibfnamefont {Y.}~\bibnamefont {Wang}}, \bibinfo
  {author} {\bibfnamefont {Y.}~\bibnamefont {Wu}}, \bibinfo {author}
  {\bibfnamefont {W.}~\bibnamefont {Hu}},\ and\ \bibinfo {author}
  {\bibfnamefont {G.}~\bibnamefont {Jin}},\ }\bibfield  {title} {\bibinfo
  {title} {{The Taiji program: A concise overview}},\ }\href
  {https://doi.org/10.1093/ptep/ptaa083} {\bibfield  {journal} {\bibinfo
  {journal} {PTEP}\ }\textbf {\bibinfo {volume} {2021}},\ \bibinfo {pages}
  {05A108} (\bibinfo {year} {2021})}\BibitemShut {NoStop}%
\bibitem [{\citenamefont {Dhurandhar}\ \emph {et~al.}(2005)\citenamefont
  {Dhurandhar}, \citenamefont {Rajesh~Nayak}, \citenamefont {Koshti},\ and\
  \citenamefont {Vinet}}]{Dhurandhar:2004rv}%
  \BibitemOpen
  \bibfield  {author} {\bibinfo {author} {\bibfnamefont {S.~V.}\ \bibnamefont
  {Dhurandhar}}, \bibinfo {author} {\bibfnamefont {K.}~\bibnamefont
  {Rajesh~Nayak}}, \bibinfo {author} {\bibfnamefont {S.}~\bibnamefont
  {Koshti}},\ and\ \bibinfo {author} {\bibfnamefont {J.~Y.}\ \bibnamefont
  {Vinet}},\ }\bibfield  {title} {\bibinfo {title} {{Fundamentals of the LISA
  stable flight formation}},\ }\href
  {https://doi.org/10.1088/0264-9381/22/3/002} {\bibfield  {journal} {\bibinfo
  {journal} {Class. Quant. Grav.}\ }\textbf {\bibinfo {volume} {22}},\ \bibinfo
  {pages} {481} (\bibinfo {year} {2005})},\ \Eprint
  {https://arxiv.org/abs/gr-qc/0410093} {arXiv:gr-qc/0410093} \BibitemShut
  {NoStop}%
\bibitem [{\citenamefont {Jiang}\ \emph {et~al.}(2025)\citenamefont {Jiang},
  \citenamefont {Xu},\ and\ \citenamefont {Zhang}}]{Jiang:2025uga}%
  \BibitemOpen
  \bibfield  {author} {\bibinfo {author} {\bibfnamefont {H.}~\bibnamefont
  {Jiang}}, \bibinfo {author} {\bibfnamefont {B.}~\bibnamefont {Xu}},\ and\
  \bibinfo {author} {\bibfnamefont {Y.-L.}\ \bibnamefont {Zhang}},\ }\bibfield
  {title} {\bibinfo {title} {{Artificial Precision Polarization Array:
  Sensitivity for the axion-like dark matter with clock satellites}},\
  }\href@noop {} {\  (\bibinfo {year} {2025})},\ \Eprint
  {https://arxiv.org/abs/2511.04400} {arXiv:2511.04400 [astro-ph.CO]}
  \BibitemShut {NoStop}%
\bibitem [{\citenamefont {Liu}\ \emph {et~al.}(2025)\citenamefont {Liu},
  \citenamefont {Zhang}, \citenamefont {Du}, \citenamefont {Liu}, \citenamefont
  {Xu},\ and\ \citenamefont {Zhang}}]{Liu:2025hwn}%
  \BibitemOpen
  \bibfield  {author} {\bibinfo {author} {\bibfnamefont {Y.-Y.}\ \bibnamefont
  {Liu}}, \bibinfo {author} {\bibfnamefont {J.-R.}\ \bibnamefont {Zhang}},
  \bibinfo {author} {\bibfnamefont {M.-H.}\ \bibnamefont {Du}}, \bibinfo
  {author} {\bibfnamefont {H.-S.}\ \bibnamefont {Liu}}, \bibinfo {author}
  {\bibfnamefont {P.}~\bibnamefont {Xu}},\ and\ \bibinfo {author}
  {\bibfnamefont {Y.-L.}\ \bibnamefont {Zhang}},\ }\bibfield  {title} {\bibinfo
  {title} {{Detectability of axion-like dark matter for different time-delay
  interferometry combinations in space-based gravitational wave detectors}},\
  }\href@noop {} {\  (\bibinfo {year} {2025})},\ \Eprint
  {https://arxiv.org/abs/2511.15438} {arXiv:2511.15438 [gr-qc]} \BibitemShut
  {NoStop}%
\end{thebibliography}%

\end{document}